\documentclass[12pt]{article}  
\usepackage{graphicx}
\usepackage{epsfig}
\usepackage{hyperref}
\usepackage {amsmath}
\usepackage {amssymb}
\usepackage {bm}  
\usepackage[usenames,dvipsnames]{color}
\usepackage{xspace} 
\newcommand{\Ncoll}{\mbox{$N_{\rm coll}$}\xspace}

\newcommand{\sqsn}{\mbox{$\sqrt{s_{_{NN}}}$}\xspace}

\def\lsim{\raise0.3ex\hbox{$<$\kern-0.75em\raise-1.1ex\hbox{$\sim$}}}
\def\gsim{\raise0.3ex\hbox{$>$\kern-0.75em\raise-1.1ex\hbox{$\sim$}}}

\def\mean#1{\left<#1\right>}
\def\Journal#1#2#3#4{{#1}{\bf #2} (#4) #3}

\def\PLB{{Phys. Lett. B}}

\def\PRC{{Phys. Rev. C}}

\def\ARNPS{{Ann. Rev. Nucl. Part. Sci.\ }}

\def\RPP{Rep. Prog. Phys.\ }

\def\QGP{{\color{Red} Q}{\color{Blue} G}{\color{Green} P}} 
\def\QCD{{\color{Red} Q}{\color{Green} C}{\color{Blue} D}} 
\normalsize
\begin{document}
\title{Latest results from RHIC + Progress on determining $\hat{q}L$ in RHI collisions using di-hadron correlations}
\author{M.~J.~Tannenbaum
\thanks{Research supported by U.~S.~Department of Energy, DE-SC0012704.}
\\Physics Department, 510c,\\
Brookhaven National Laboratory,\\
Upton, NY 11973-5000, USA\\
mjt@bnl.gov} 
\date{}
\maketitle
\thispagestyle{empty} 
\vspace*{-2pc}
\begin{abstract}
Results from Relativistic Heavy Ion Collider Physics in 2018 and plans for the future at Brookhaven National Laboratory are presented.
\end{abstract}

\section{Introduction}\label{sec:introduction}\vspace*{-1.0pc}
\begin{figure}[!h]
\begin{center}
\raisebox{0pc}{\includegraphics[width=0.90\textwidth]{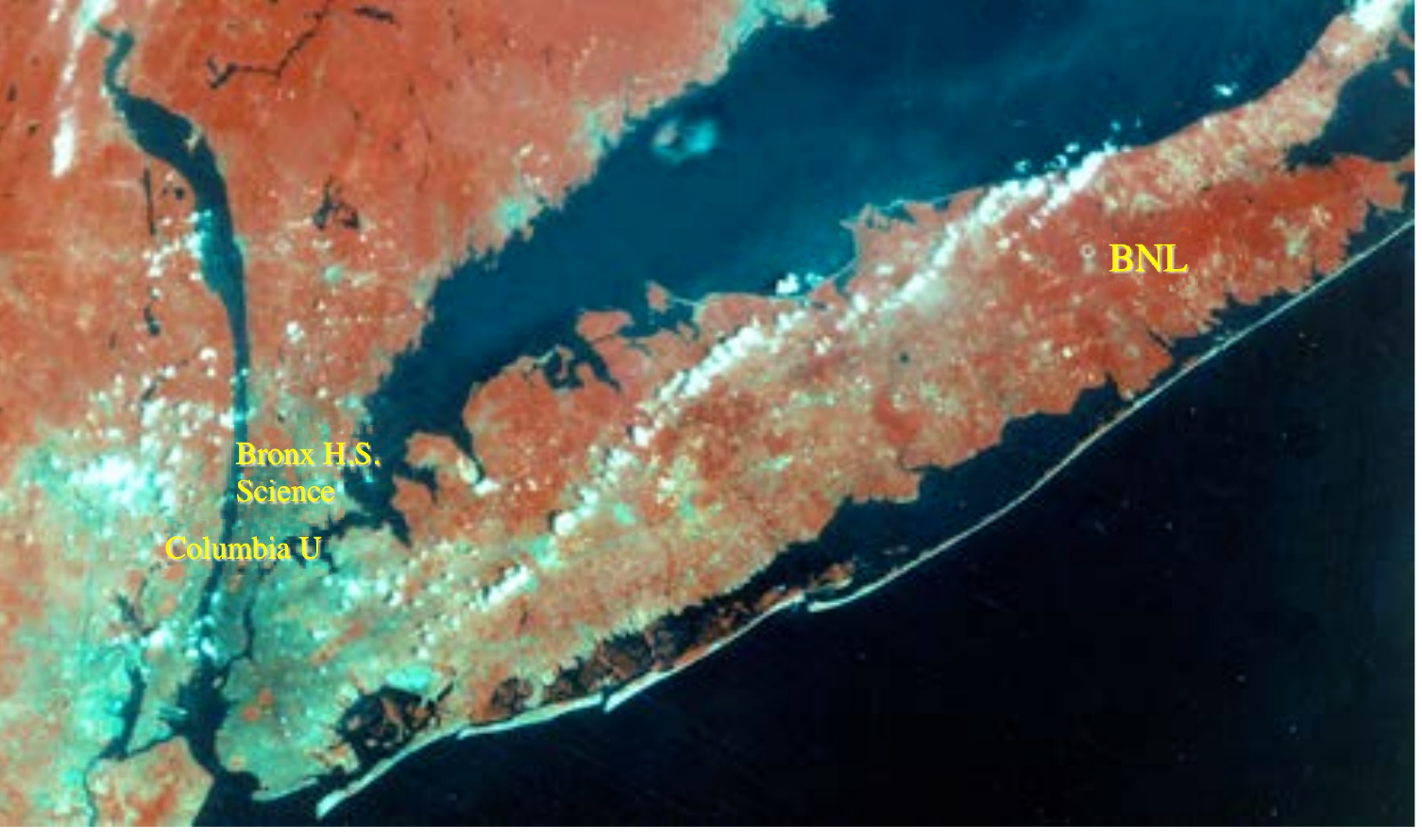}}\hspace*{0.2pc}
\end{center}\vspace*{-1.5pc}
\caption[]{\footnotesize NASA infra-red photo of Long Island and the New York Metro Region from space. RHIC is the white circle to the left of the word BNL. Manhattan Island in New York City, $\sim$100 km west of BNL, is also clearly visible on the left side of the photo, with Columbia U. and Bronx Science High School indicated.}
\label{fig:BNLphoto}\vspace*{-0.5pc}
\end{figure}

\begin{figure}[!h]
\begin{center}
\raisebox{0pc}{\includegraphics[height=0.54\textwidth,width=0.75\textwidth]{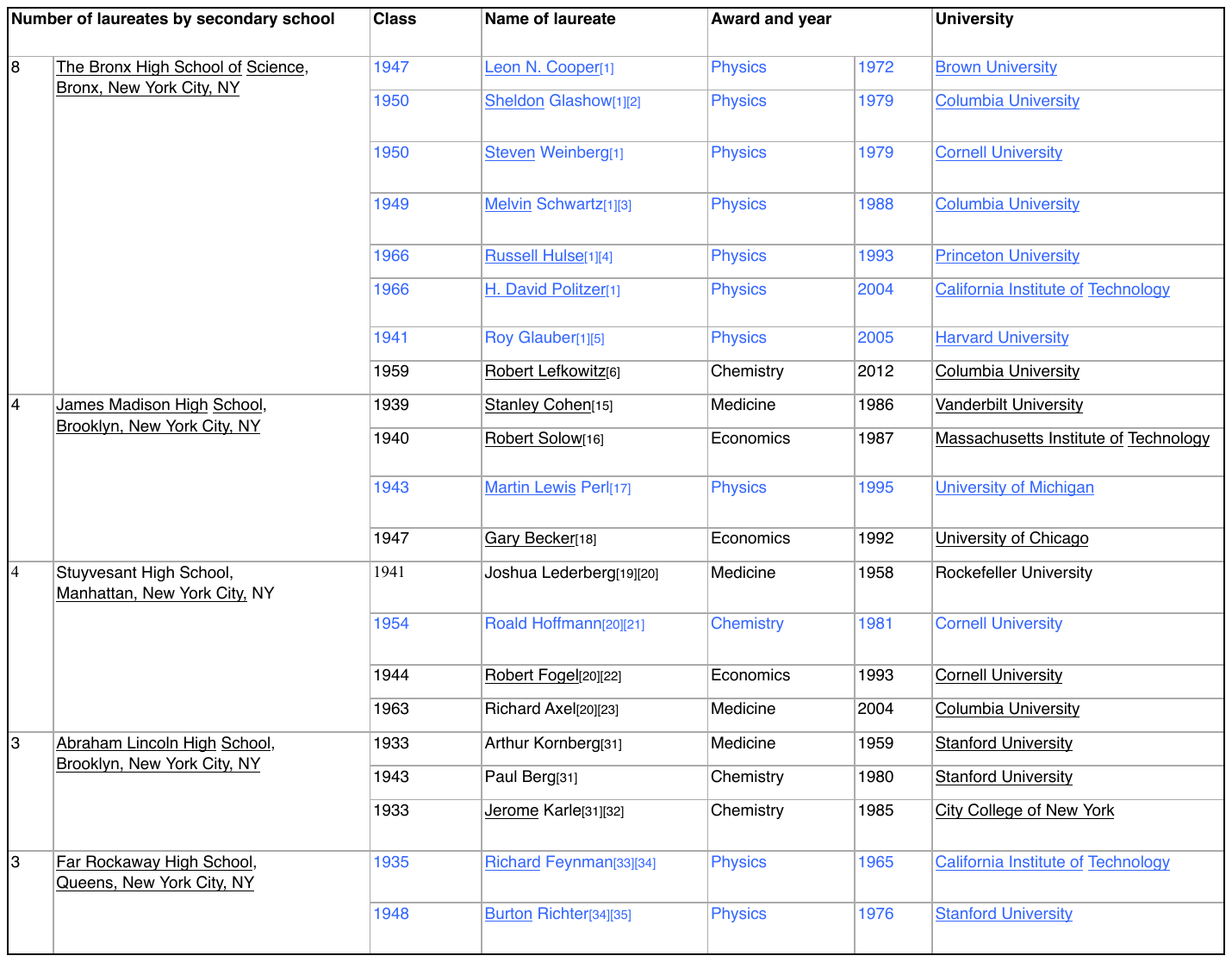}}\\
\raisebox{0pc}{\includegraphics[height=0.54\textwidth,width=0.75\textwidth]{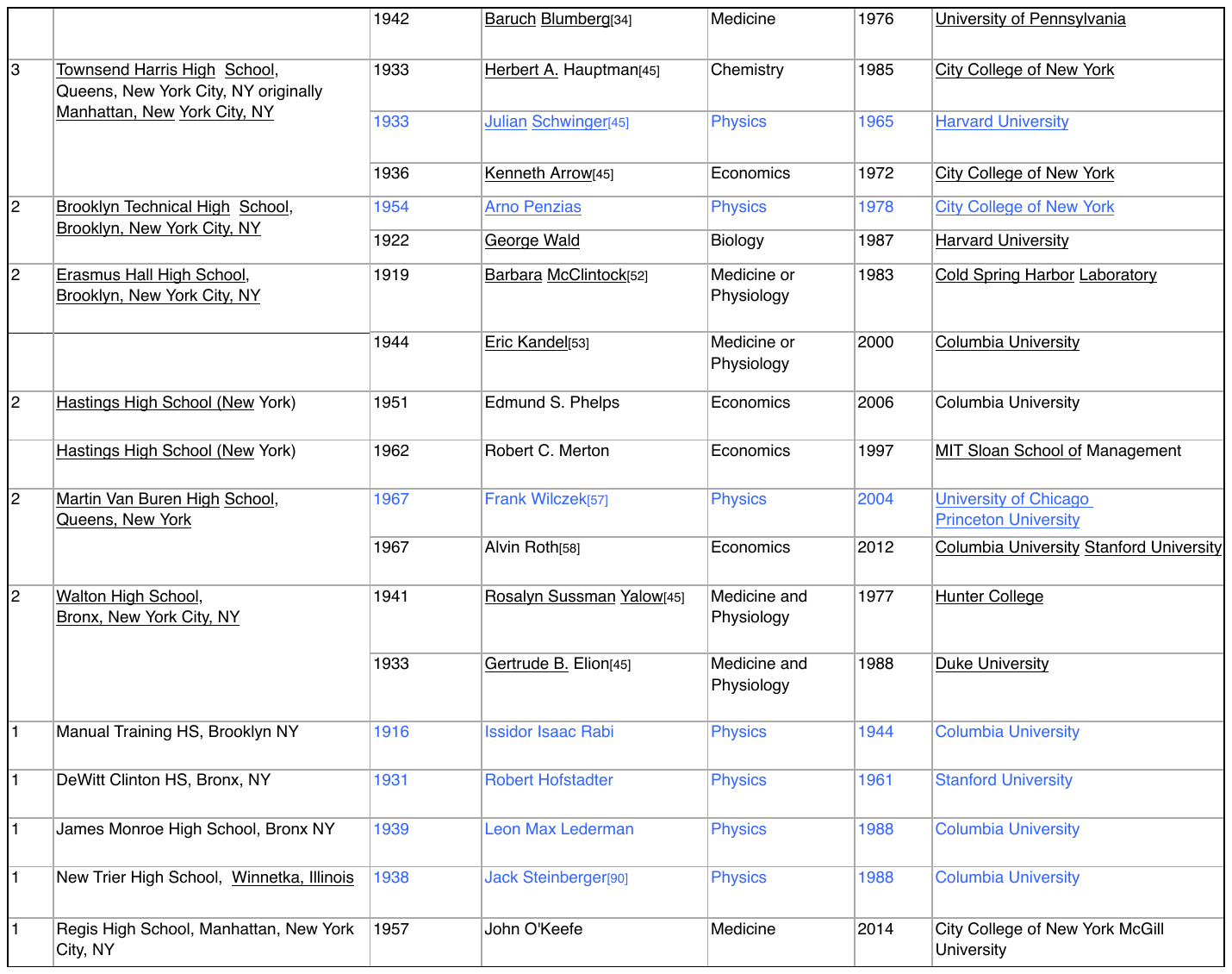}}\vspace*{-1.5pc}
\end{center}
\caption[]{\footnotesize From Wikipedia (edited), Physicists in {\color{blue} blue} +  
Roald Hoffman a classmate of mine from Columbia.}
\label{fig:NYCHSnobel}\vspace*{-0.5pc}
\end{figure}

The Relativistic Heavy Ion Collider (RHIC) at Brookhaven National Laboratory (BNL) is one of the two remaining operating hadron colliders in the world, and the first and only polarized p$+$p collider. BNL is located in the center of the roughly 200 km long maximum 40 km wide island (named Long Island), and appears on the map as the white circle which is the berm containing the Relativistic Heavy Ion Collider (RHIC).  BNL is 100 km from New York City in a region which nurtures science with Columbia University and the Bronx High School of Science indicated (Fig.~\ref{fig:BNLphoto}). Perhaps more convincing is the list of the many Nobel Prize winners from New York City High School graduates (Fig.~\ref{fig:NYCHSnobel}) which does not yet include one of this years Nobel Prize winners in Physics, Arthur Ashkin who graduated from James Madison High school in 1940 and Columbia U. in 1947.   

\begin{figure}[!h]
\begin{center}
\raisebox{0pc}{\includegraphics[width=0.90\textwidth]{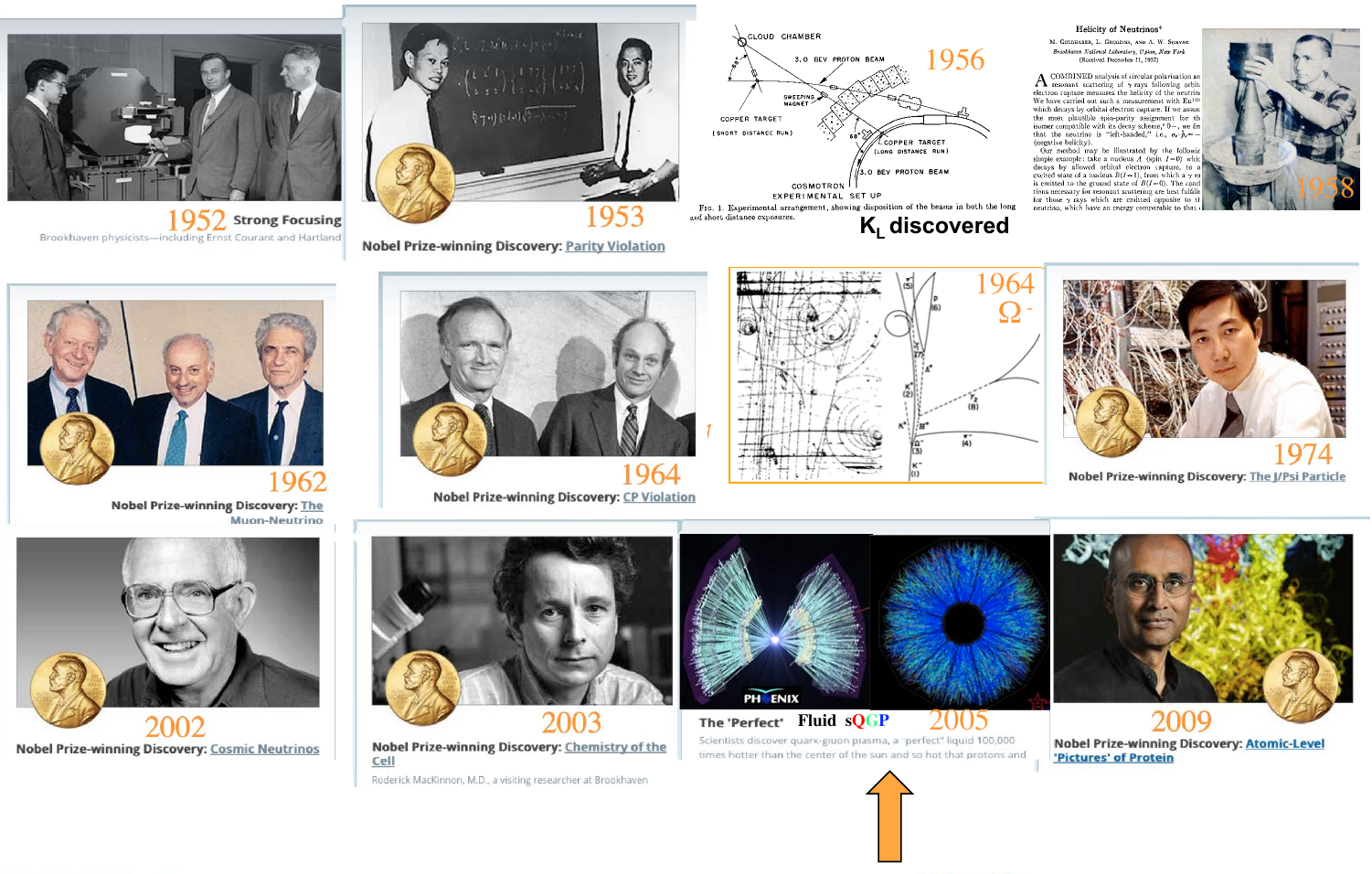}}\hspace*{0.2pc}
\end{center}\vspace*{-2.0pc}
\caption[]{\footnotesize Selected Discoveries and Nobel Prizes at BNL, arrow points to \QGP\ discovery.}
\label{fig:BNLdiscoveries}\vspace*{-1.0pc}. 
\end{figure}

There also have been many discoveries and Nobel Prizes at BNL (Fig.~\ref{fig:BNLdiscoveries}).  

In particular, Leon Lederman who made many discoveries at BNL died this past year (2018) at the age of 96. Leon was the most creative and productive high energy physics experimentalist of his generation as well as the physicist with the best jokes. He was also my PhD thesis Professor. For more details see\\  
\url{https://physicstoday.scitation.org/do/10.1063/PT.6.4.20181010a/full/}
\begin{figure}[!h]
\begin{center}
\raisebox{0pc}{\includegraphics[width=0.90\textwidth]{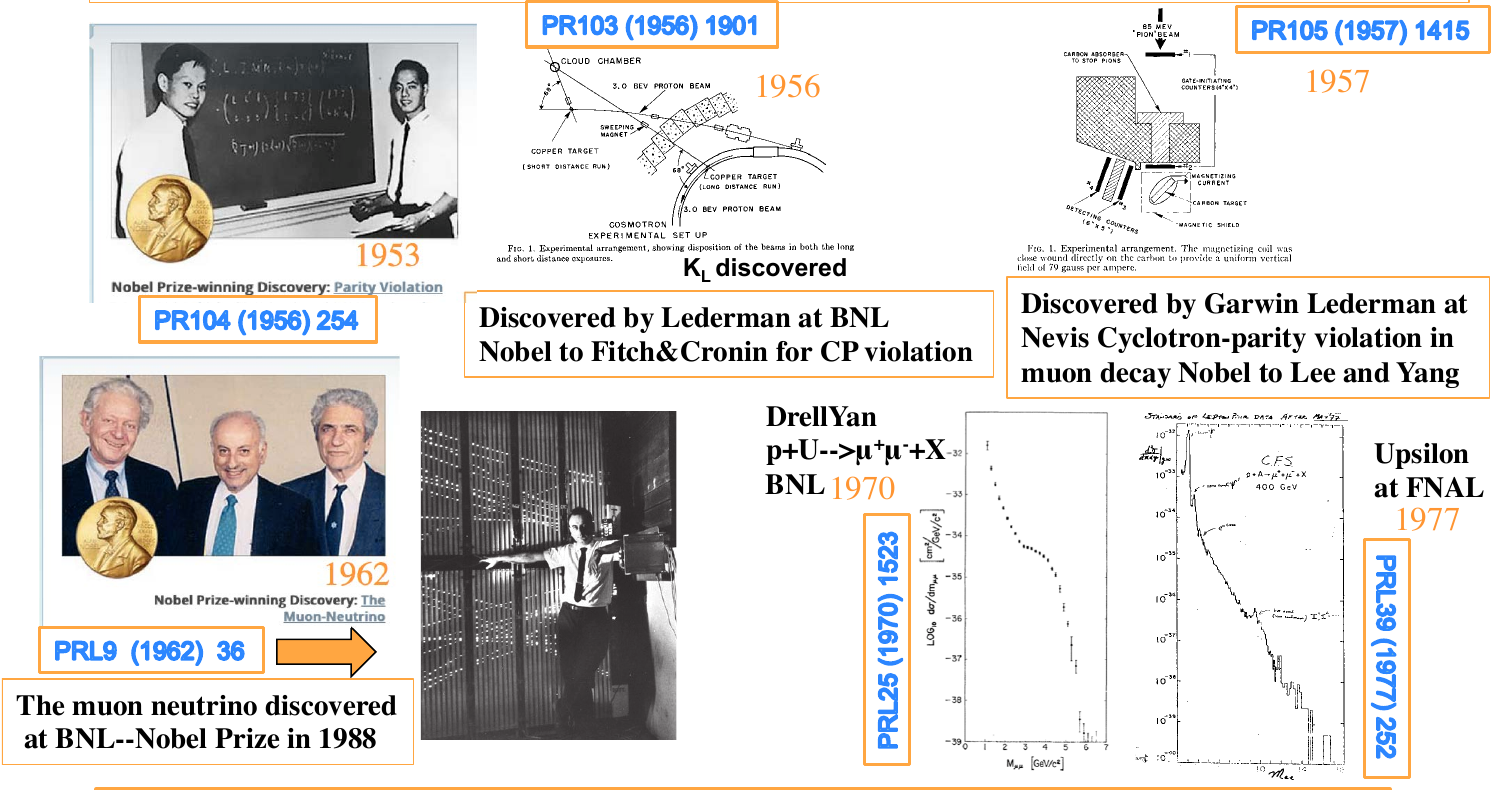}}\hspace*{0.2pc}
\end{center}\vspace*{-1.5pc}
\caption[]{\footnotesize Discoveries by Leon Lederman and close associates at Columbia University.}
\label{fig:LMLBNLdiscoveries}\vspace*{-0.5pc}
\end{figure}

\section{Why RHIC was built: to discover the \QGP. }
\begin{figure}[!h]
\begin{center}
\raisebox{0pc}{\includegraphics[width=0.80\textwidth]{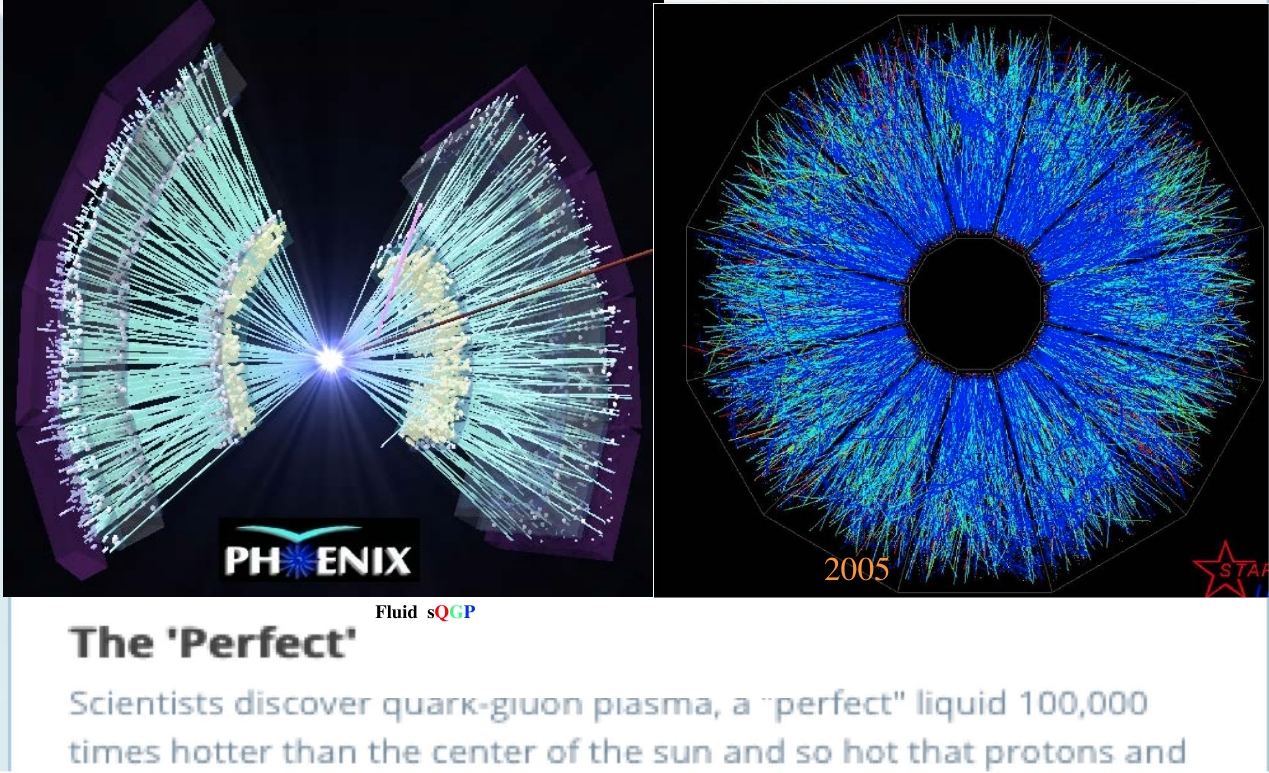}}\hspace*{0.2pc}
\end{center}\vspace*{-1.5pc}
\caption[]{\footnotesize View along the beam direction of central collision events in Au$+$Au collisions in the PHENIX and STAR detectors at RHIC.}
\label{fig:PXstar}\vspace*{-0.5pc}
\end{figure}

Figure~\ref{fig:PXstar} shows central collision particle production in the PHENIX and STAR detectors, which were the major detectors at RHIC. 

\begin{figure}[!h]
\begin{center}
\raisebox{0pc}{\includegraphics[width=0.80\textwidth]{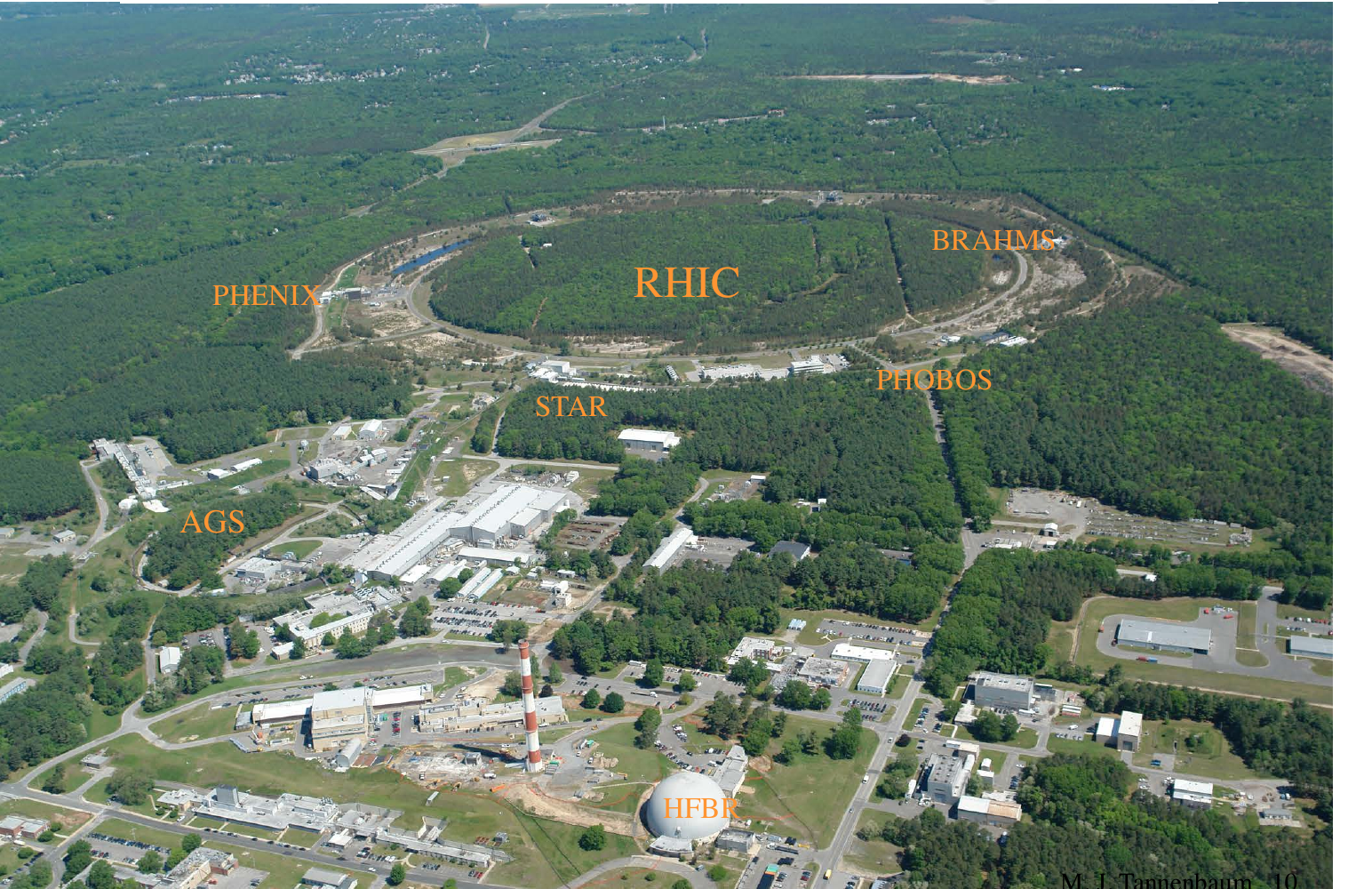}}\hspace*{0.2pc}
\end{center}\vspace*{-1.5pc}
\caption[]{\footnotesize View of RHIC location from the air. The positions of the 4 original detectors, PHENIX, STAR PHOBOS and BRAHMS are indicated as well as the AGS (with 3 Nobel Prizes shown in Fig.~\ref{fig:BNLdiscoveries}).}
\label{fig:BNLRHICcloseup}\vspace*{-0.5pc} 
\end{figure}

At the startup of RHIC in the year 2000 there were two smaller more special purpose detectors PHOBOS and BRAHMS as shown in Fig.~\ref{fig:BNLRHICcloseup}, which finished data taking in 2005. 
\subsection{The first major RHIC experiments}
\begin{figure}[!h]
\begin{center}
\raisebox{0pc}{\includegraphics[width=0.90\textwidth]{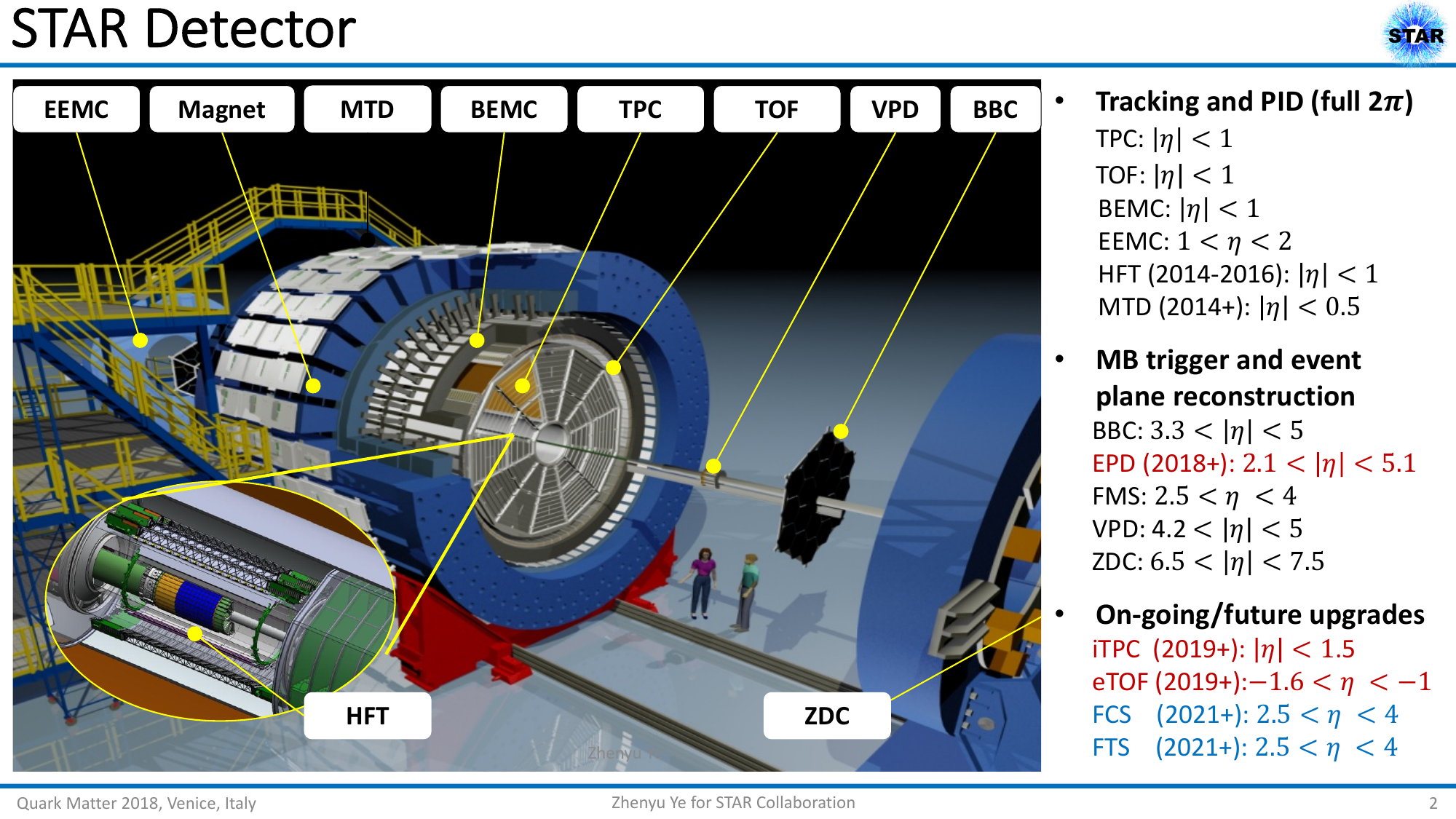}}\hspace*{0.2pc}
\end{center}\vspace*{-1.5pc}
\caption[]{\footnotesize STAR is based on a normal conductor solenoid with Time Projection Chamber for tracking, an EM Calorimeter, Vertex detector and $\mu$ detector behind the thick iron yoke. }
\label{fig:STAR2018}\vspace*{-0.5pc}
\end{figure}

\begin{figure}[!h]
\begin{center}
\raisebox{0pc}{\includegraphics[width=0.90\textwidth]{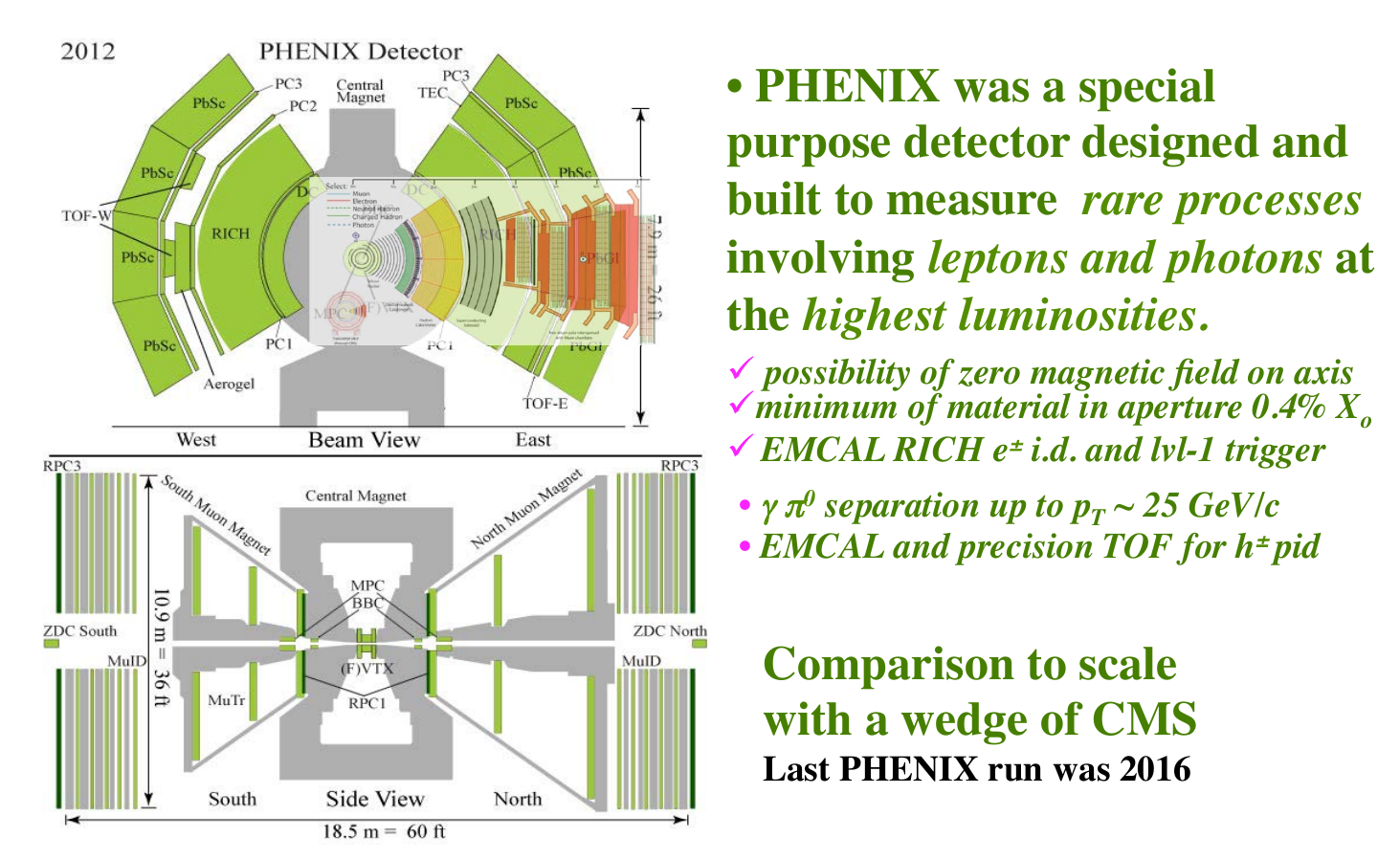}}\hspace*{0.2pc}
\end{center}\vspace*{-1.5pc}
\caption[]{\footnotesize As indicated on the figure, PHENIX is a special perpose detector for electrons and photons but also measures charged hadrons and notably $\pi^0\rightarrow \gamma+\gamma$ at mid-rapidity and muons in the forward direction.}
\label{fig:PHENIX2002}\vspace*{-0.5pc}
\end{figure}

The two major experiments at RHIC were STAR (Fig.~\ref{fig:STAR2018}), which is still operating, and PHENIX (Fig.~\ref{fig:PHENIX2002}) which finished data taking at the end of the 2016 run. 

\subsection{The new major RHIC experiment sPHENIX}
\begin{figure}[!h]
\begin{center}
\raisebox{0pc}{\includegraphics[width=0.90\textwidth]{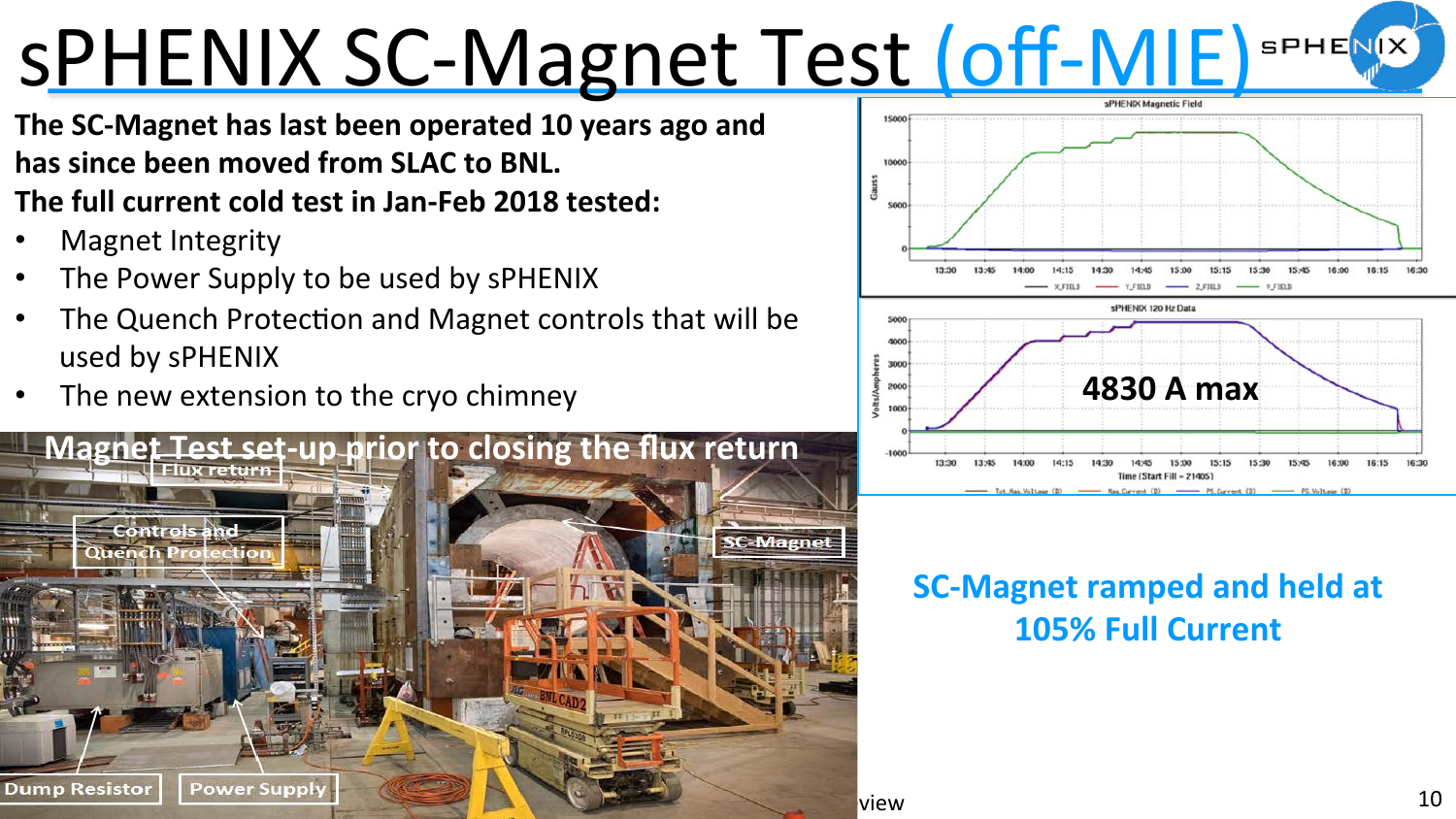}}\hspace*{0.2pc}
\end{center}\vspace*{-1.5pc}
\caption[]{\footnotesize BABAR superconducting solenoid now in operation at BNL}
\label{fig:sPmagnet}\vspace*{-0.5pc}
\end{figure}

sPHENIX is a major improvement over PHENIX with a superconducting thin coil solenoid which was surplus from the BABAR experiment at SLAC and is now working at BNL and has reached its full field (Fig.~\ref{fig:sPmagnet}).

\begin{figure}[!h]
\begin{center}
\raisebox{0pc}{\includegraphics[width=0.90\textwidth]{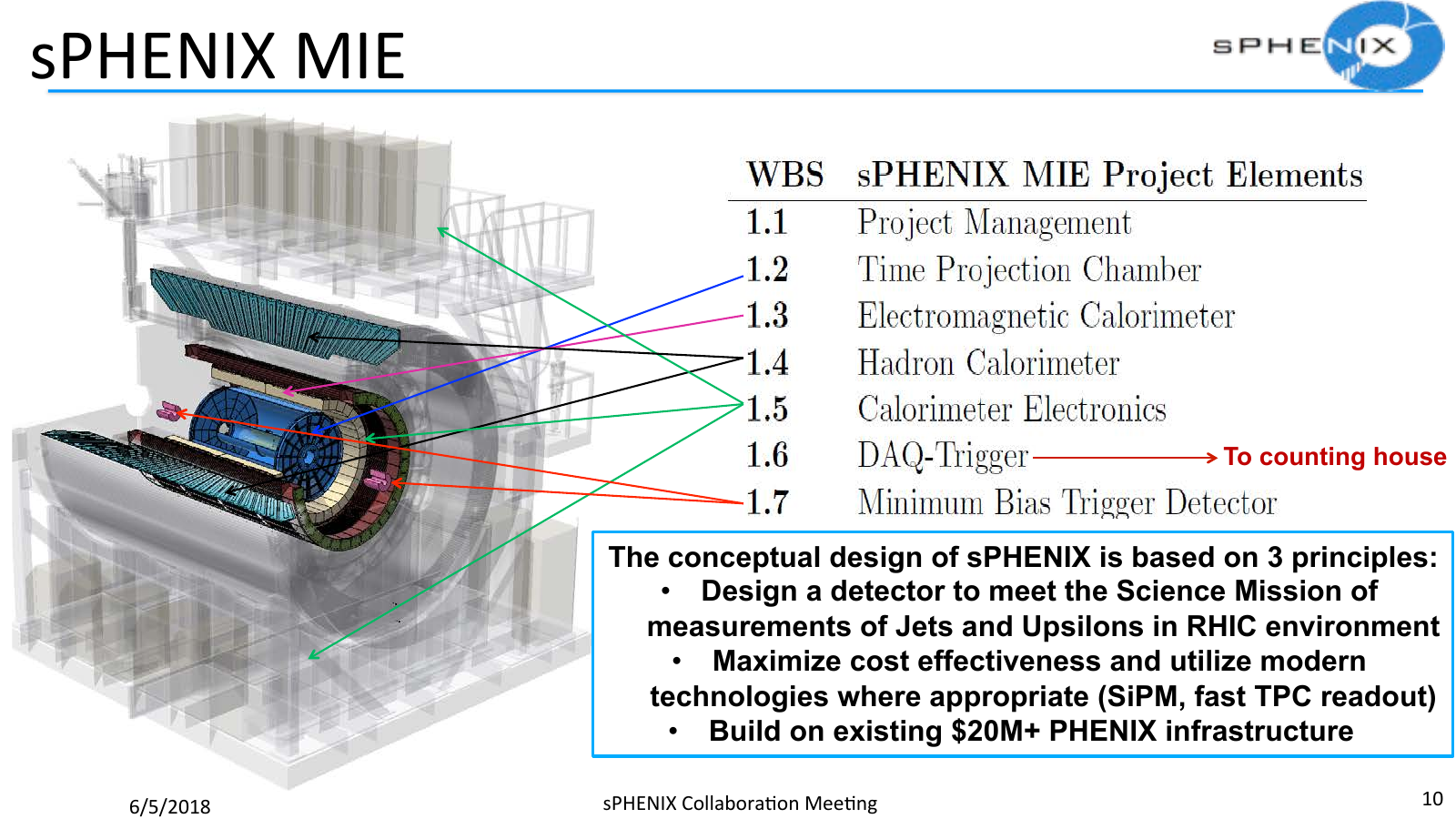}}\hspace*{0.2pc}
\end{center}\vspace*{-1.5pc}
\caption[]{\footnotesize Conceptual design of sPHENIX with major features illustrated.}
\label{fig:sPHENIX}\vspace*{-0.5pc}
\end{figure}

The design of the sPHENIX experiment is moving along well (Fig.~\ref{fig:sPHENIX}) with a notable addition of a hadron calorimeter based on the iron return yoke of the solenoid. 

\begin{figure}[!h]
\begin{center}
\raisebox{0pc}{a)\includegraphics[width=0.90\textwidth]{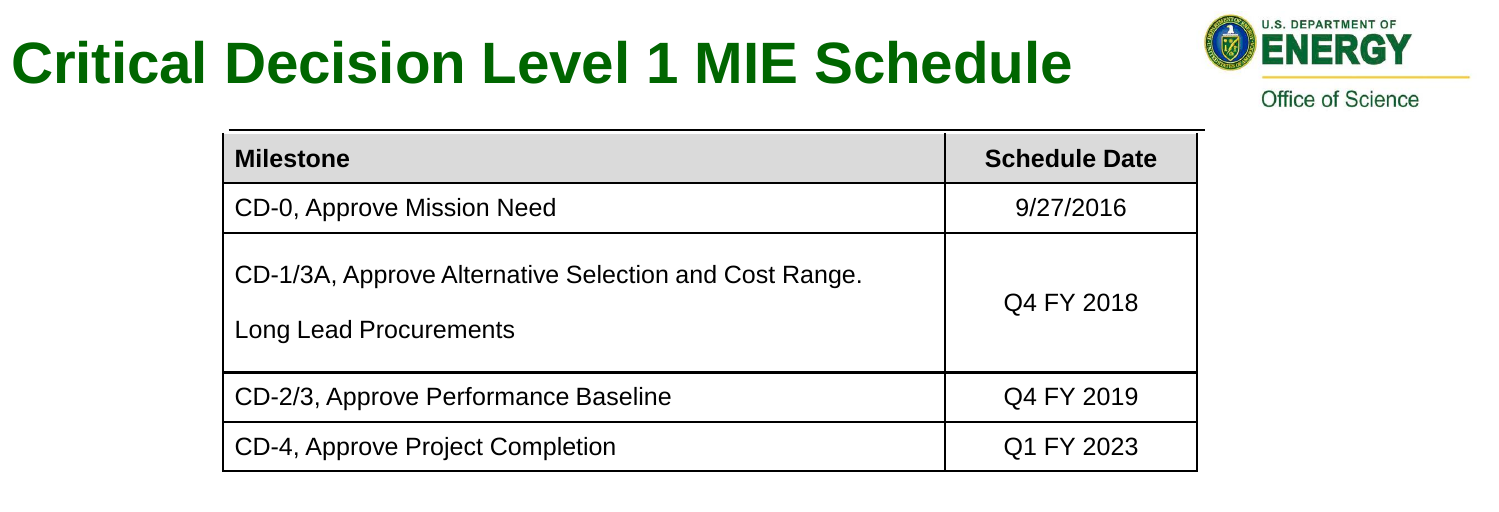}}\hspace*{0.2pc}\\
\raisebox{0pc}{b)\includegraphics[width=0.90\textwidth]{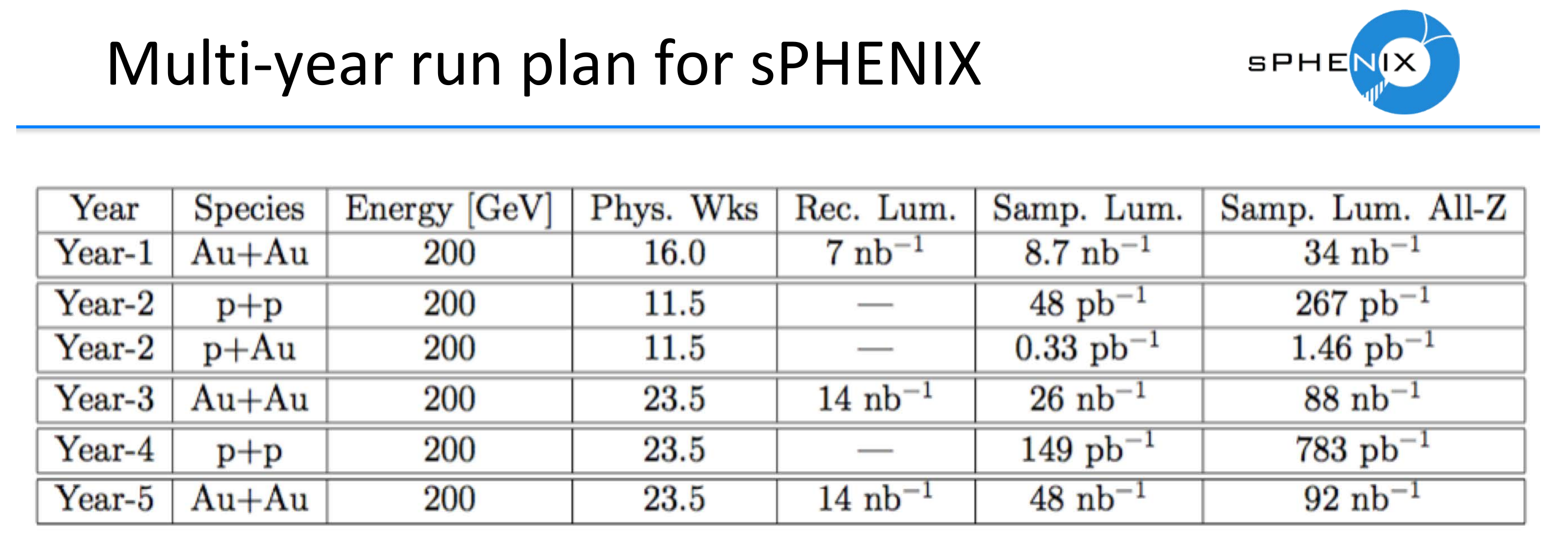}}\hspace*{0.2pc}
\end{center}\vspace*{-1.5pc}
\caption[]{\footnotesize a) DoE Critical Decision Schedule and b) Multi-year run plan for sPHENIX. }
\label{fig:DoECD1}\vspace*{-0.5pc}
\end{figure}
\begin{figure}[!h]
\begin{center}
\raisebox{0pc}{\includegraphics[width=0.90\textwidth]{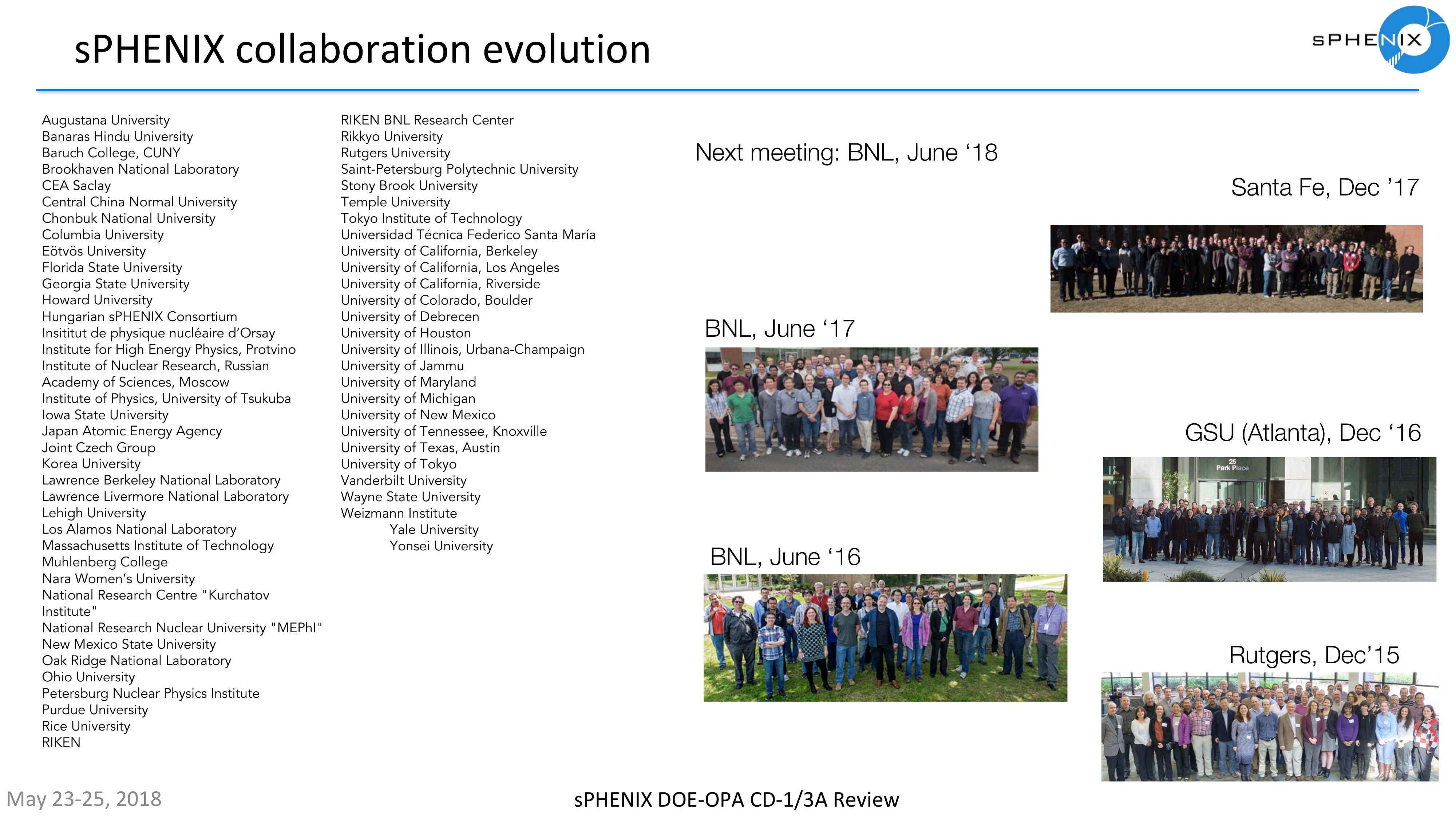}}\hspace*{0.2pc}
\end{center}\vspace*{-1.5pc}
\caption[]{\footnotesize List of the sPHENIX collaboration members in June 2018 together with photos showing the evolution since December 2015. Dave Morrison (BNL) and Gunther Roland (MIT) are spokespersons.}
\label{fig:sPHENIXcollab}\vspace*{-1.5pc}
\end{figure}

sPHENIX has been approved by the U.~S. Department of Energy (DoE) as a Major Item of Equipment(MIE) with the schedule of critical decisions shown in Fig.~\ref{fig:DoECD1}a, and the planned multi-year RHIC runs indicated in Fig.~\ref{fig:DoECD1}b. The present sPHENIX collaboration and its evolution is shown in Fig.~\ref{fig:sPHENIXcollab}. \vspace*{-0.5pc}

\subsection{Following RHIC in U.S. Nuclear Physics: the EIC. }\vspace*{-0.8pc}
\begin{figure}[!h]
\begin{center}
\raisebox{0pc}{\includegraphics[width=0.95\textwidth]{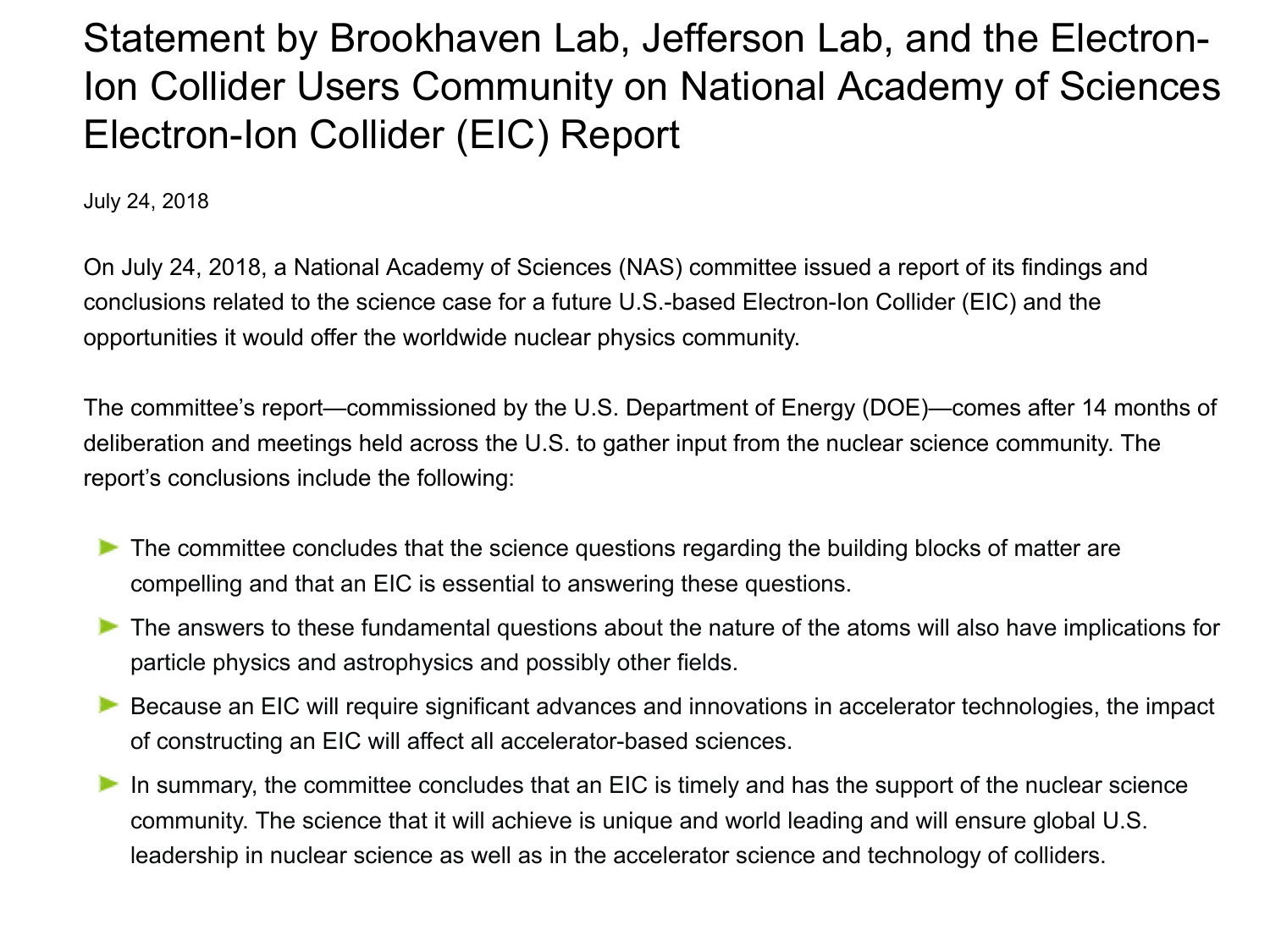}}\hspace*{0.2pc}
\end{center}\vspace*{-1.5pc}
\label{fig:NASEIC}\vspace*{-1.5pc}
\end{figure}

\begin{figure}[!h]
\begin{center}
\raisebox{0pc}{\includegraphics[width=0.78\textwidth]{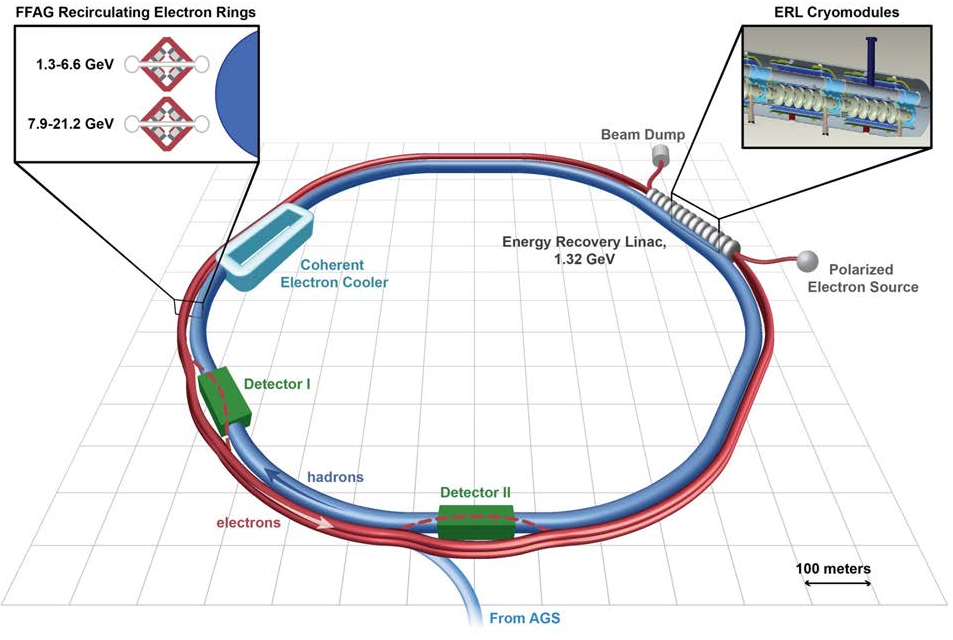}}\hspace*{0.2pc}
\end{center}\vspace*{-1.5pc}
\caption[]{\footnotesize 2014 Cost estimate: BNL \$755.9M; Temple NSAC subcommittee cost estimate \$1.5B}
\label{fig:BNLEIC2014}\vspace*{-0.5pc}
\end{figure}

\begin{figure}[!h]
\begin{center}
\raisebox{0pc}{\includegraphics[width=0.99\textwidth]{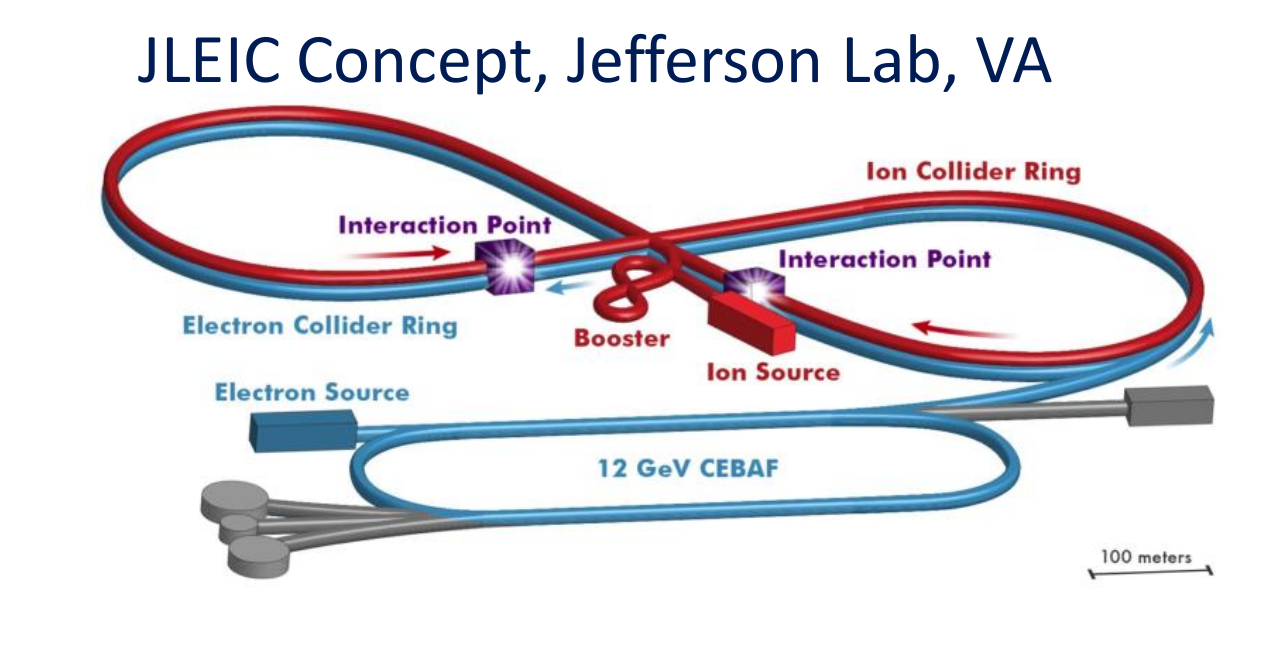}}\hspace*{0.2pc}
\end{center}\vspace*{-1.5pc}
\caption[]{\footnotesize JLab EIC Concept. Temple committee cost estimate also \$1.5B 
but no new accelerator technology required}
\label{fig:JLABEIC}\vspace*{0.5pc}
\end{figure}

\begin{figure}[!h]
\begin{center}
\raisebox{0pc}{\includegraphics[width=0.99\textwidth]{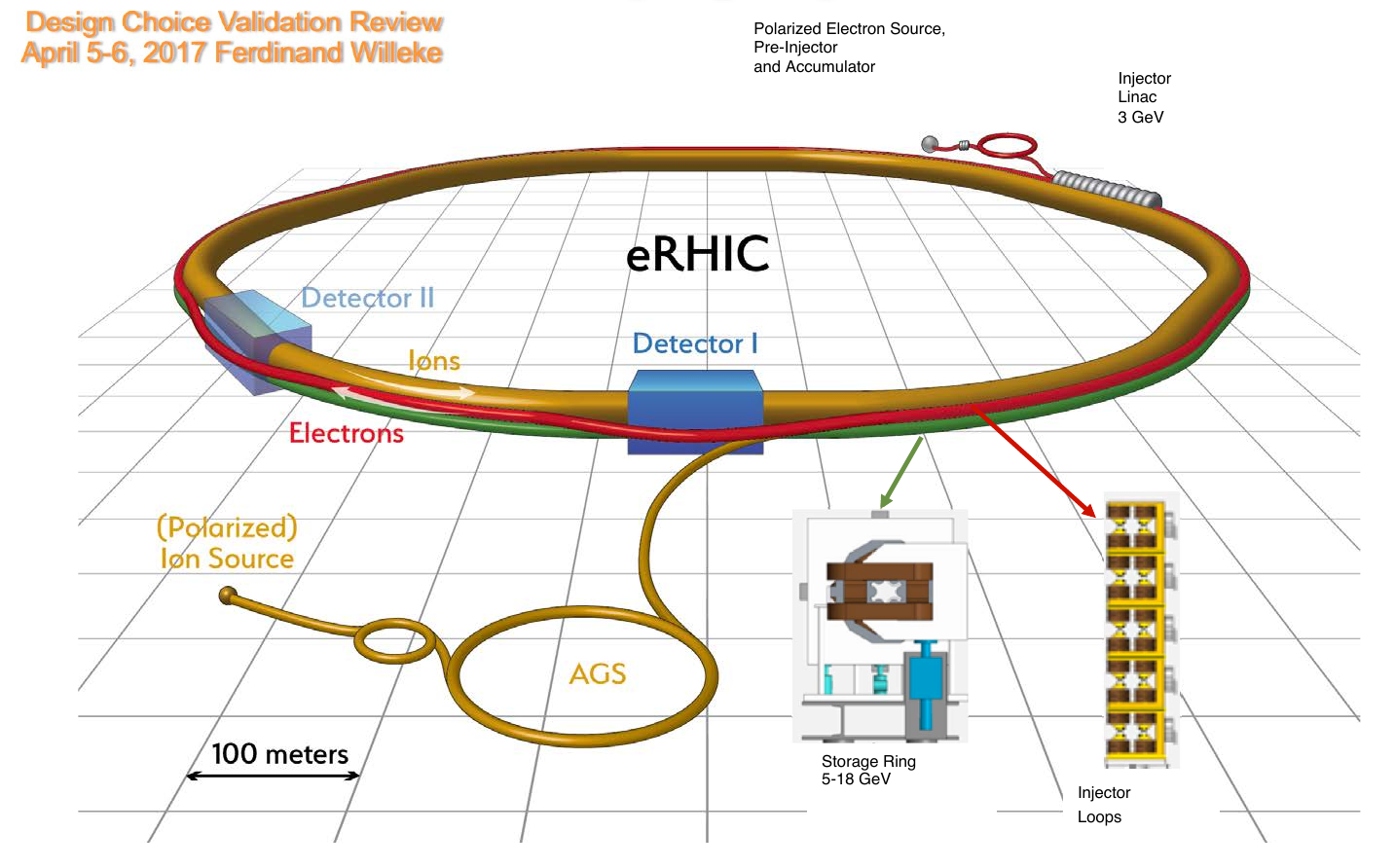}}\hspace*{0.2pc}
\end{center}\vspace*{-1.5pc}
\caption[]{\footnotesize BNL eRHIC design progress 2017. Temple committee cost estimate \$1.5B}
\label{fig:BNLEIC}
\end{figure}

The first BNL EIC design in 2014 is shown in Fig.~\ref{fig:BNLEIC2014}. The 2018 JLab and BNL EIC designs are shown in Figs.~\ref{fig:JLABEIC},\ref{fig:BNLEIC}.

The two new designs of the JLab (JLEIC) and BNL (eRHIC) both satisfy the Temple committee cost estimate of \$1.5B, but R\&D of the novel first BNL design is not idle. 

\subsubsection{R\&D for an improved less expensive BNL machine is ongoing}
\begin{figure}[!h]
\begin{center}
\raisebox{0pc}{a)\includegraphics[width=0.90\textwidth]{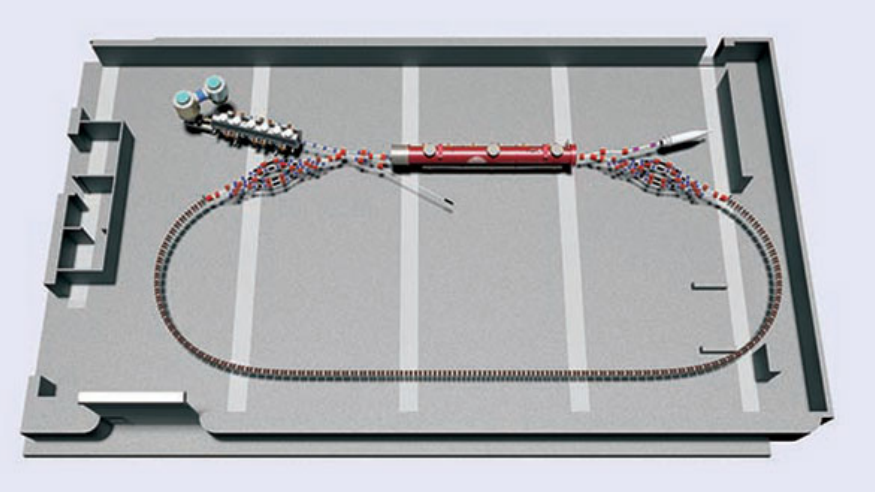}}\hspace*{0.2pc}\\
b)\raisebox{0pc}{\includegraphics[width=0.45\textwidth]{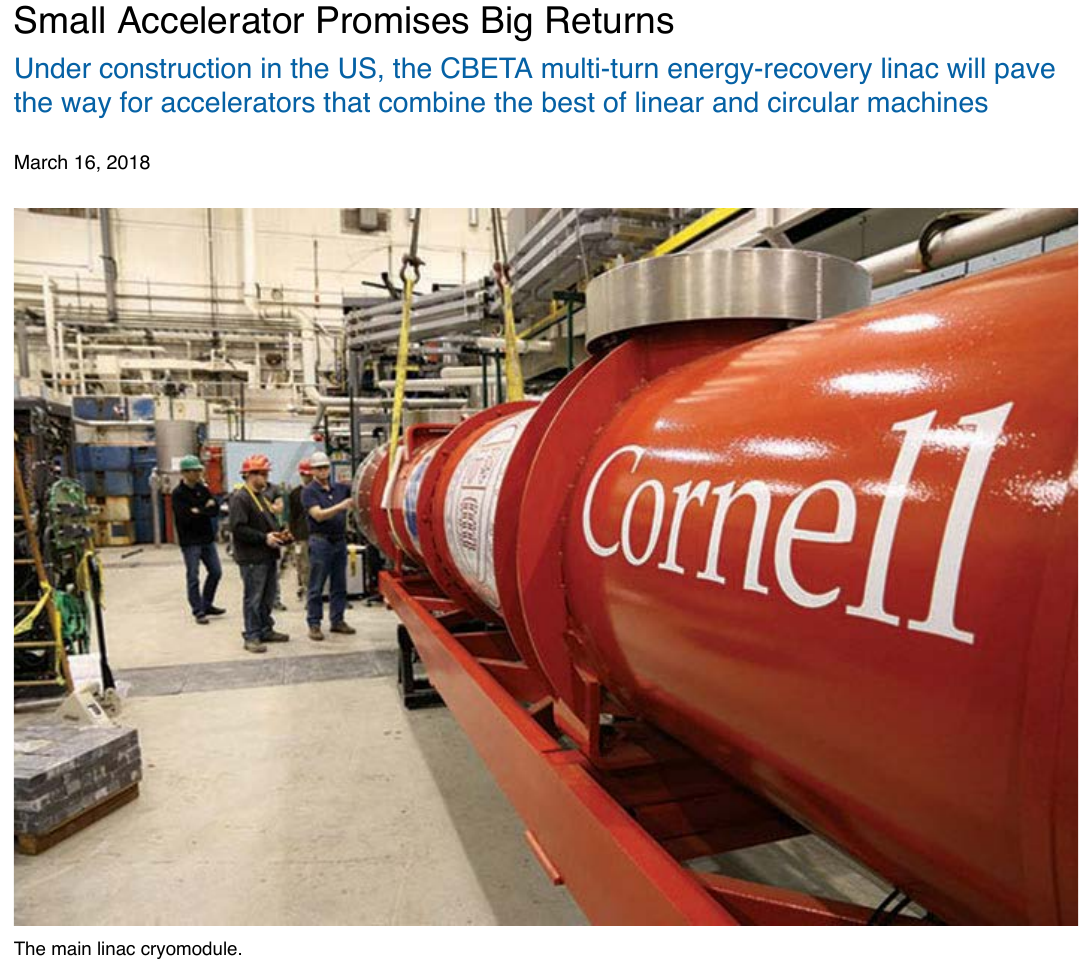}}\hspace*{0.2pc}
c)\raisebox{0pc}{\includegraphics[width=0.45\textwidth]{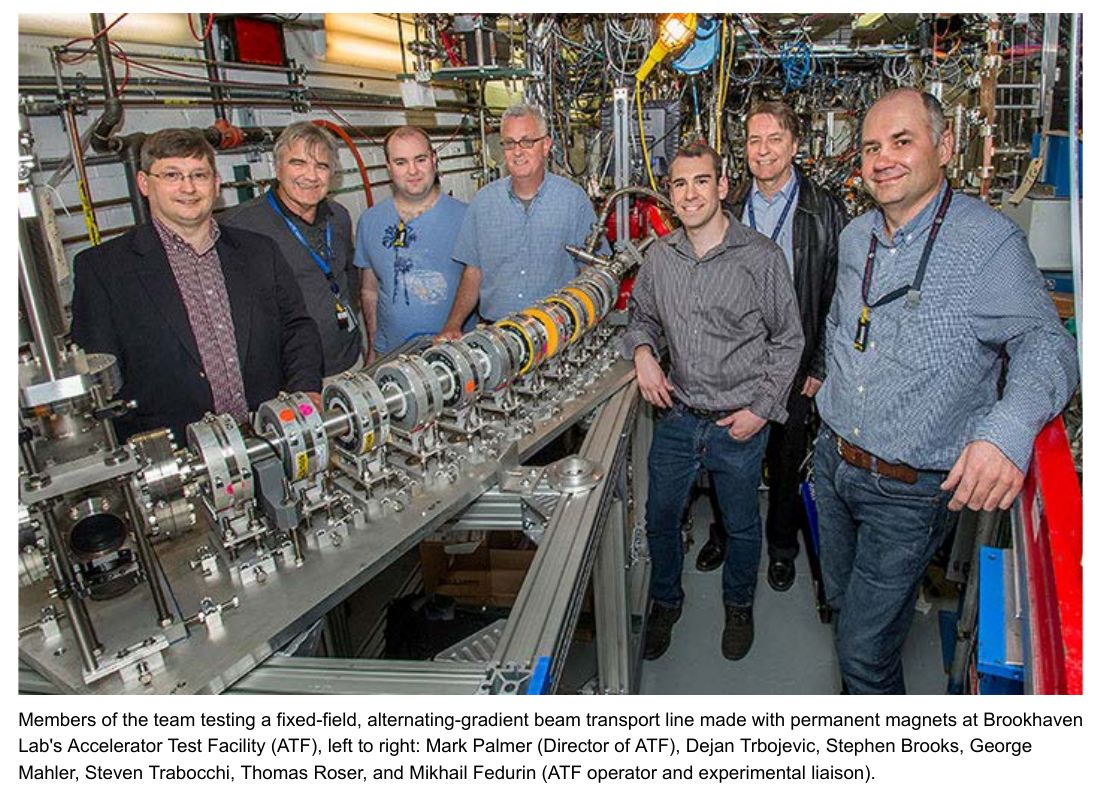}}
\end{center}\vspace*{-1.5pc}
\caption[]{\footnotesize a) CBETA (Cornell-Brookhaven Energy Recovery Linac (ERL))  b) Main Linac cryo module c) FFAG permanent loop return loop.}
\label{fig:BNLprogress}\vspace*{-0.5pc}
\end{figure}

BNL and Cornell are in the process of experiments studying an energy recovery linac ERL (Fig. \ref{fig:BNLprogress}a). Fig. \ref{fig:BNLprogress}b is the main Linac cryo module made from superconducting RF cavities. Fig. \ref{fig:BNLprogress}c is a return loop made from fixed-field alternating-gradient (FFAG) optics made with permanent Halbach magnets to contain four beam energies in a single 70 mm-wide beam pipe, designed and prototyped at Brookhaven National Laboratory (BNL).

\pagebreak\newpage
\section{\large  RHIC future Run Plan and and the present RHIC run in 2018}\vspace*{-1.5pc}
\begin{figure}[!h]
\begin{center}
\raisebox{0pc}{\includegraphics[width=0.85\textwidth]{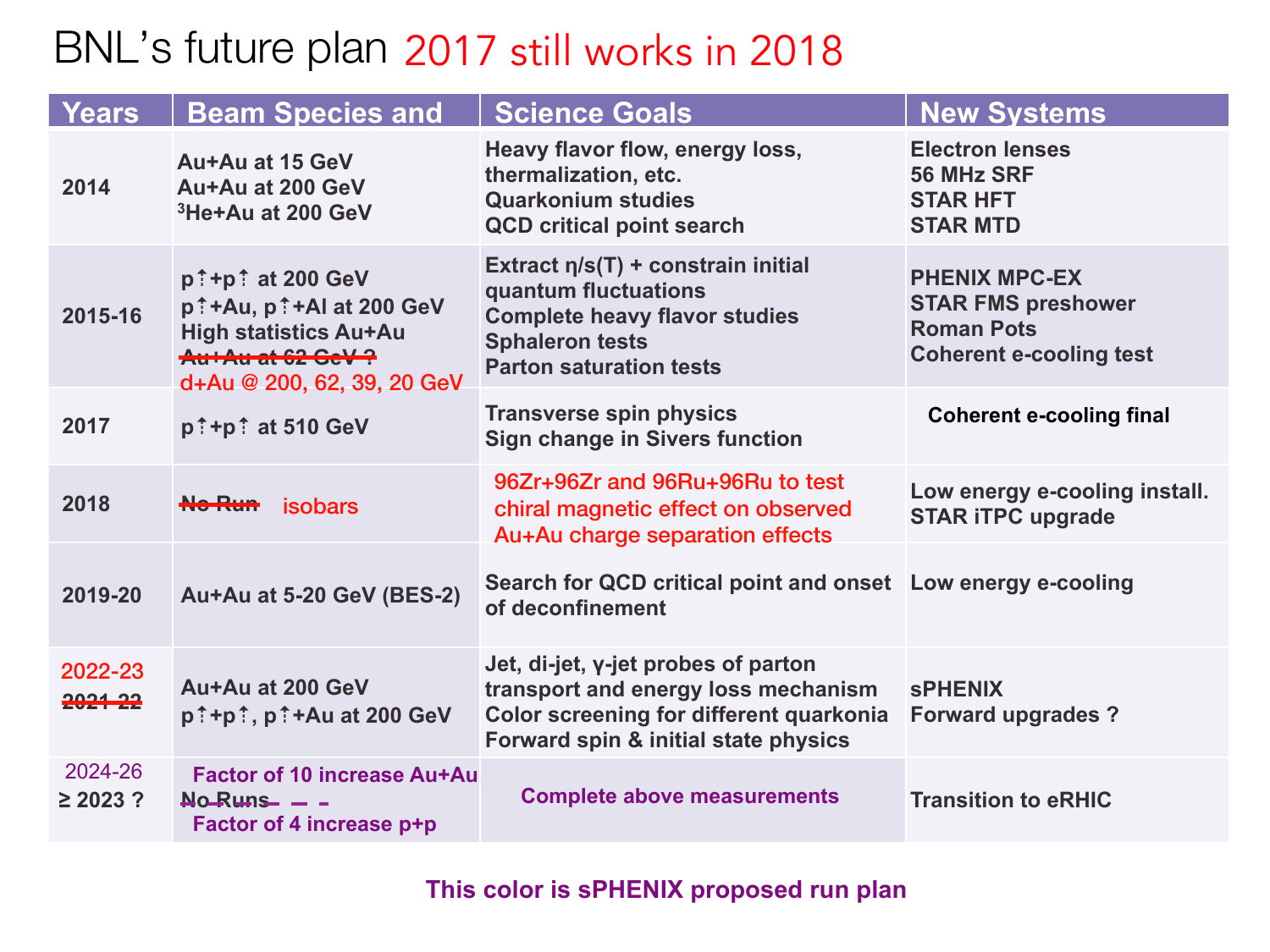}}\hspace*{0.2pc}\\
\end{center}\vspace*{-2.5pc}
\caption[]{\footnotesize RHIC run plan 2014-2023 (2026?).  }
\label{fig:2018RHICrunplan}\vspace*{-0.5pc}
\end{figure}
\begin{figure}[!h]
\begin{center}
\raisebox{0pc}{\includegraphics[width=0.85\textwidth]{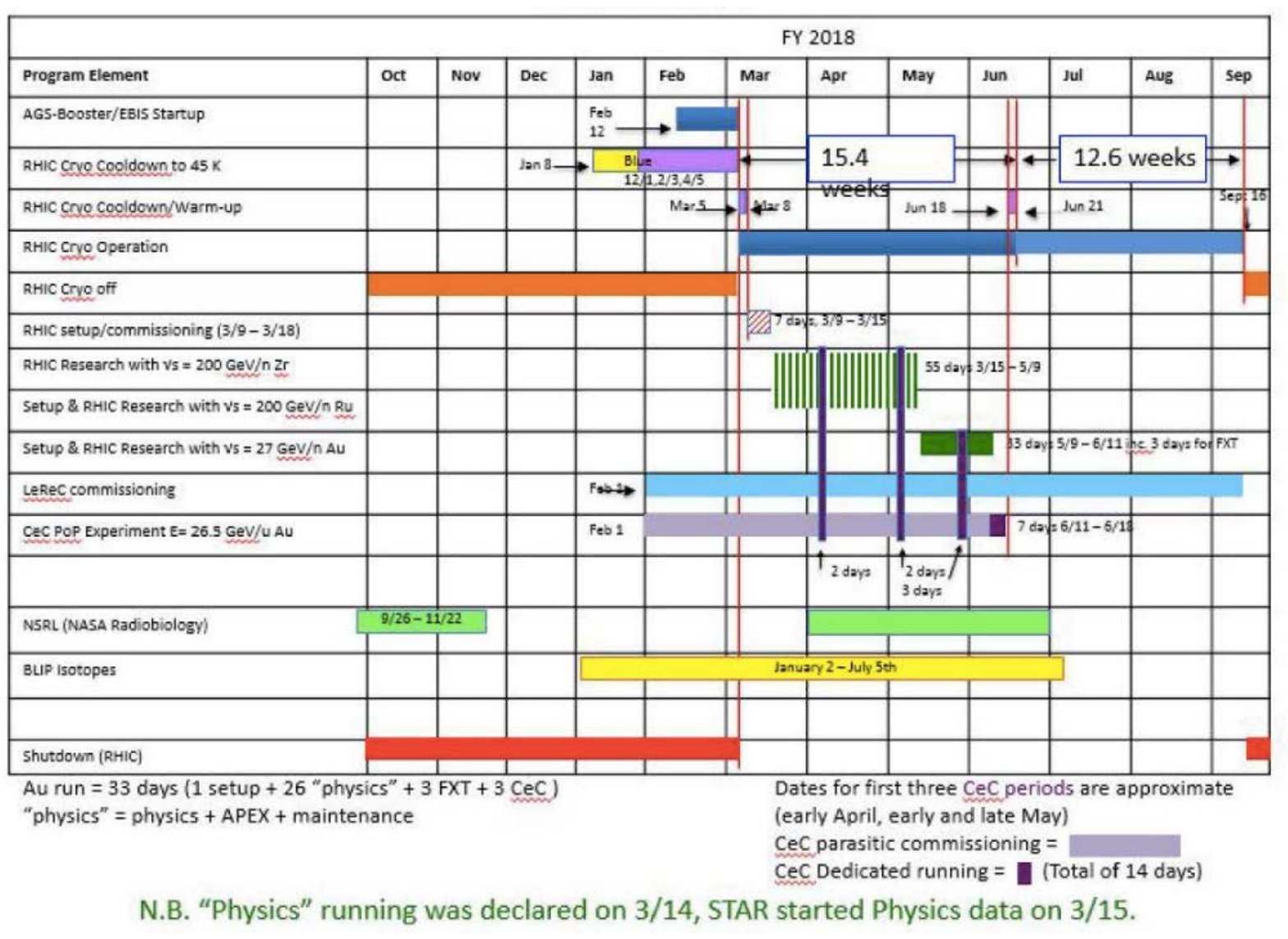}}\hspace*{0.2pc}
\end{center}\vspace*{-2.5pc}
\caption[]{\footnotesize  2018 RHIC Run Schedule. }
\label{fig:2018RHICrun}\vspace*{-1.5pc}
\end{figure}

\subsection{2018 RHIC run is $_{40}$Zr$^{96}$ + $_{40}$Zr$^{96}$ and $_{44}$Ru$^{96}$ + $_{44}$Ru$^{96}$ , why?}
\begin{figure}[!h]
\begin{center}
\raisebox{0pc}{\includegraphics[width=0.99\textwidth]{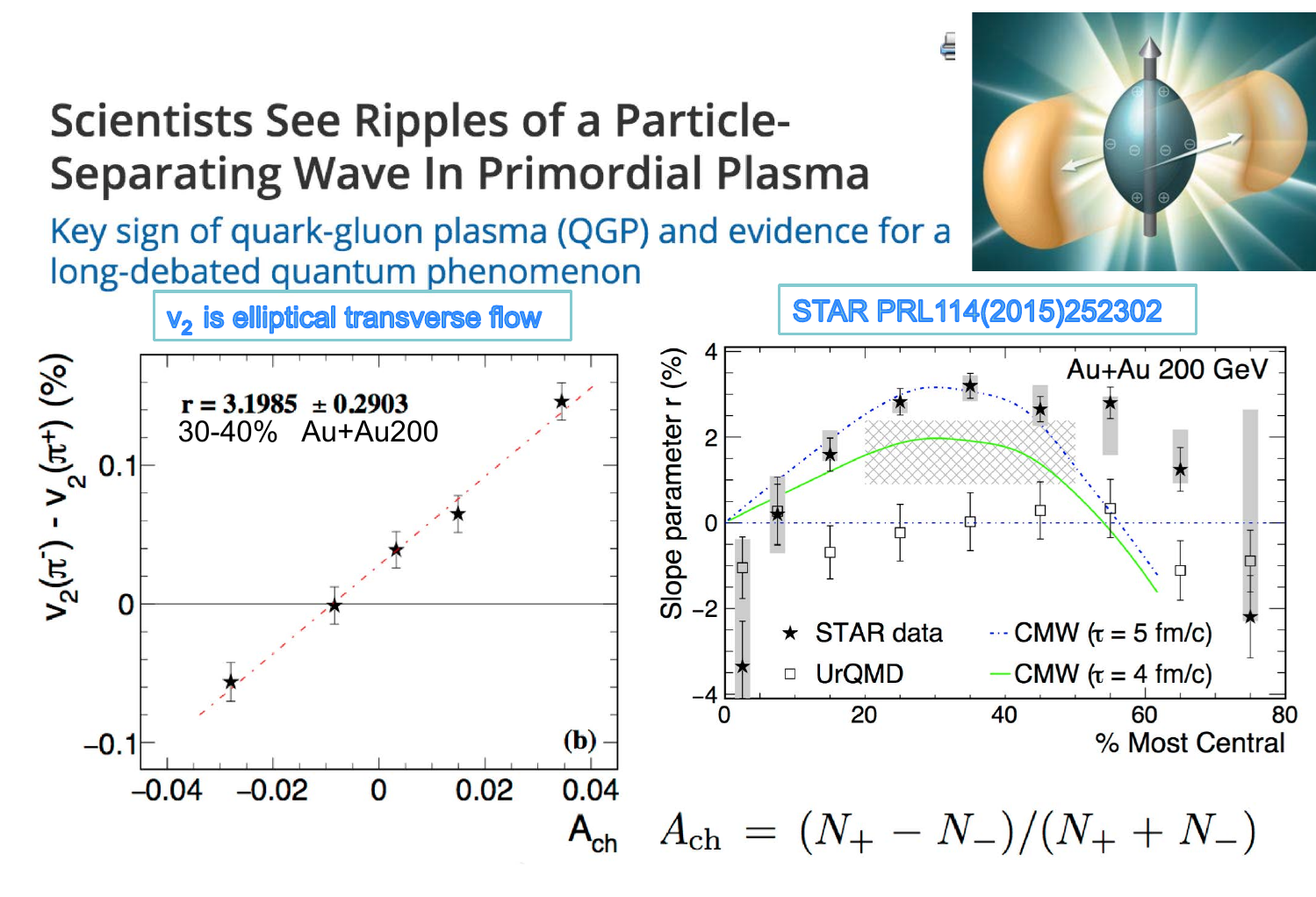}}\hspace*{0.2pc}
\end{center}\vspace*{-2.5pc}
\caption[]{\footnotesize From article by Karen McNulty Walsh in BNL news June 8,2015}
\label{fig:2015RHICrun}\vspace*{0.5pc}
\end{figure}

In order to determine whether the separation of charges in the flow, $v_2$, of $\pi^+$ and $\pi^-$ shown in Fig.~\ref{fig:2015RHICrun} is due to a new phenomenon called the Chiral Magnetic Effect (Fig.~\ref{fig:2018RHICrun}a) the 2018 measurements are made with collisions of Zr$+$Zr and Ru$+$Ru which have the same number of nucleons but different electric charges (Fig.~\ref{fig:2018RHICrun}b). If the effect is larger in Ru$+$Ru with stronger charge and magnetic field compared to Zr$+$Zr with the same number of nucleons, it will indicate that the charge asymmetry is the Chiral Magnetic Effect.  
\begin{figure}[!h]
\begin{center}
\raisebox{0pc}{a)\includegraphics[width=0.28\textwidth]{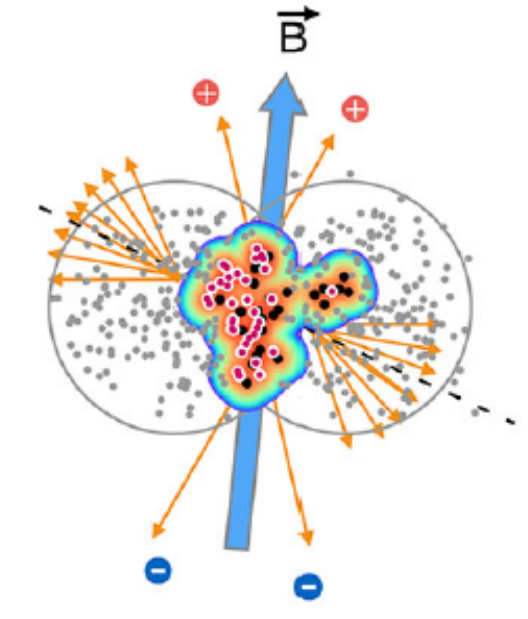}}\hspace*{0.2pc}
\raisebox{0pc}{b)\includegraphics[width=0.63\textwidth]{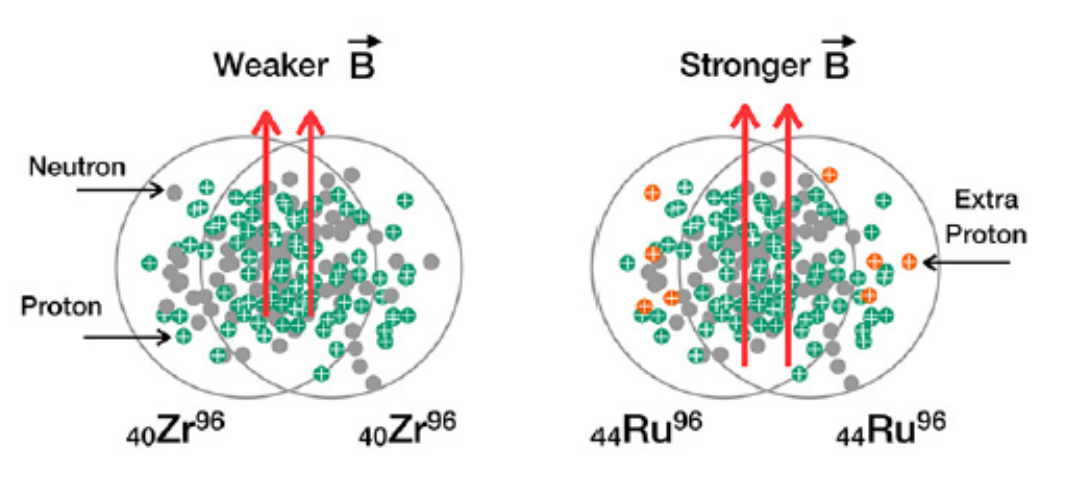}}\hspace*{0.2pc}
\end{center}\vspace*{-1.5pc}
\caption[]{\footnotesize a) schematic of A+A collision. b) sketch of the stronger magnetic (B) field in Ru+Ru.}
\label{fig:2018RHICrun}\vspace*{-0.5pc}
\end{figure}

\subsection{Vorticity: an application of particle physics to the \QGP }
\begin{figure}[!h]
\begin{center}
a)\raisebox{1pc}{\includegraphics[width=0.55\textwidth]{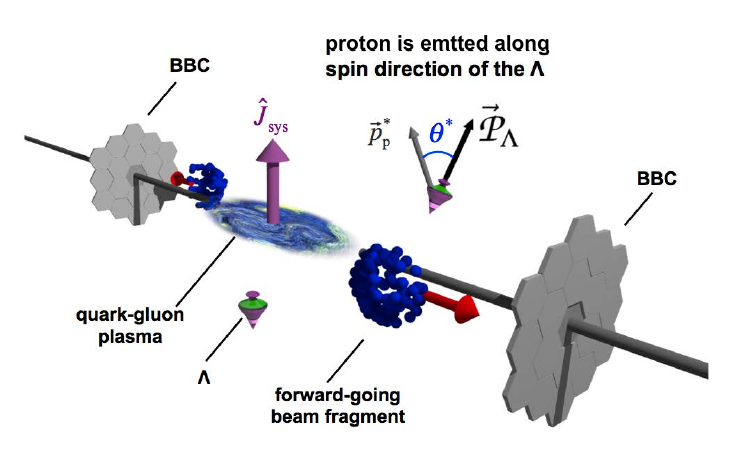}}\hspace*{-1.2pc}
\raisebox{0pc}{b)\includegraphics[width=0.34\textwidth]{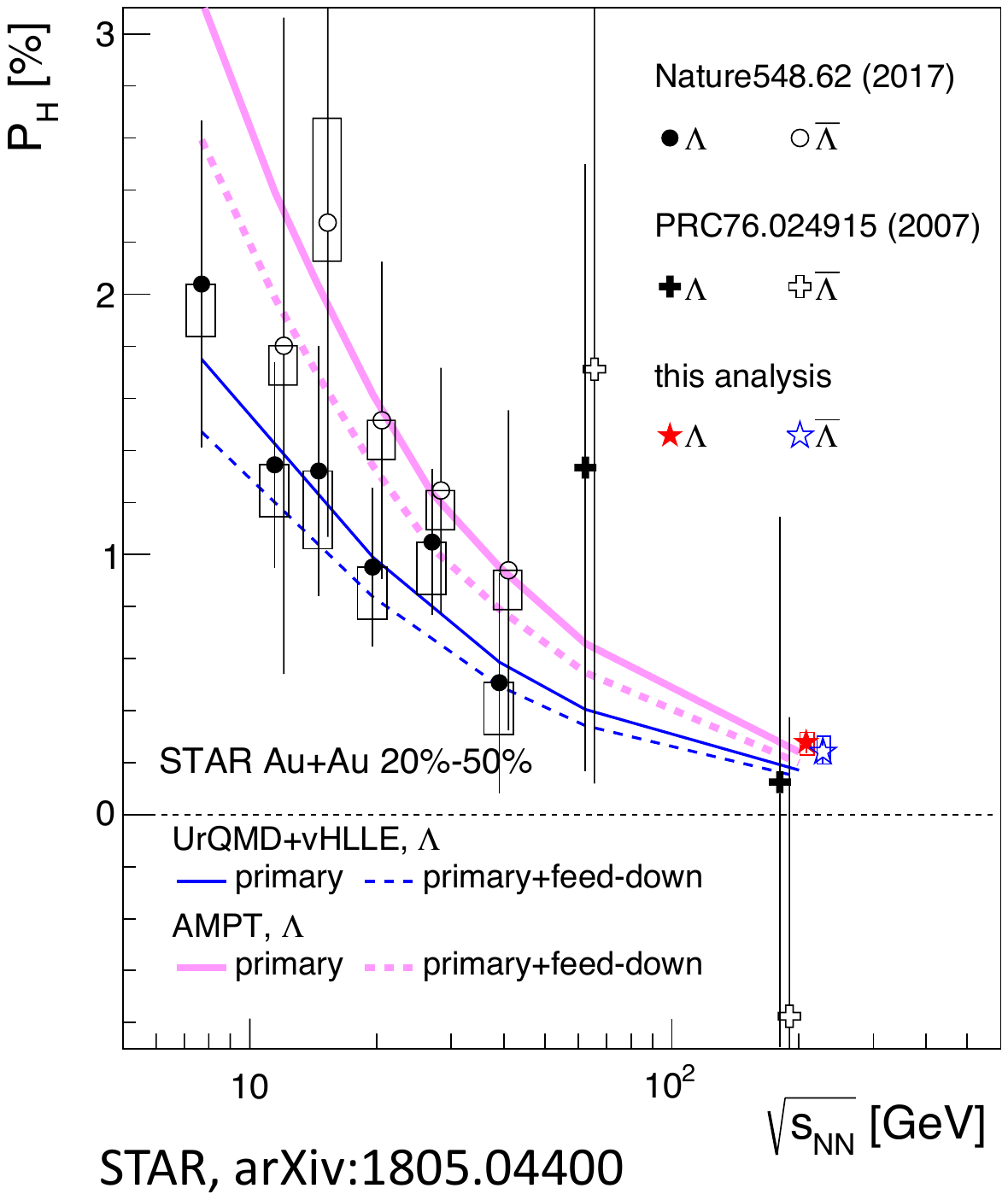}}\hspace*{0.2pc}
\end{center}\vspace*{-1.5pc}
\caption[]{\footnotesize a) Schematic of STAR vorticity detection. b) Polarization P$_{\rm H}$=$\overline{{\cal P}}_\Lambda$ or  $\overline{{\cal P}}_{\overline \Lambda}$ vs \sqsn }
\label{fig:vorticity}\vspace*{-0.5pc}
\end{figure}

It was observed at FERMILAB {[\small \Journal{PRL\ }{36}{1113}{1976}]} that forward $\Lambda$ were polarized in p$+$Be collisions, where the proton in the $\Lambda\rightarrow p+\pi^-$ decay is emitted along the spin direction of the $\Lambda$. In the A+A collision (Fig.~\ref{fig:vorticity}a), the forward going beam fragments are deflected outwards so that the event plane and the angular momentum $\hat{J}_{sys}$ of the \QGP\ formed can be determined. STAR claims that the $\Lambda$ polarization, $\overline{{\cal P}}_\Lambda$, is parallel to the angular momentum  $\hat{J}_{sys}$ of the \QGP\ everywhere so that the vorticity $\omega=k_B T (\overline{{\cal P}}_\Lambda + \overline{{\cal P}}_{\overline \Lambda})/\hbar$ can be calculated, a good exercise for the reader to see if you can get the $\omega\sim 10^{22}/s$ which is $10^5$ times larger than any other fluid [\Journal{Nature\ }{548}{62-65}{2017}]. Another interesting thing to note is that the largest vorticity is at $\sqsn=7.6-19$ GeV where the CERN fixed target experiments measure. Does this mean that their fluid (with minimal if any \QGP) is also perfect?!!!\vspace{-2.0pc}  

\begin{figure}[!h]
\begin{center}
\raisebox{0pc}{\includegraphics[width=0.60\textwidth]{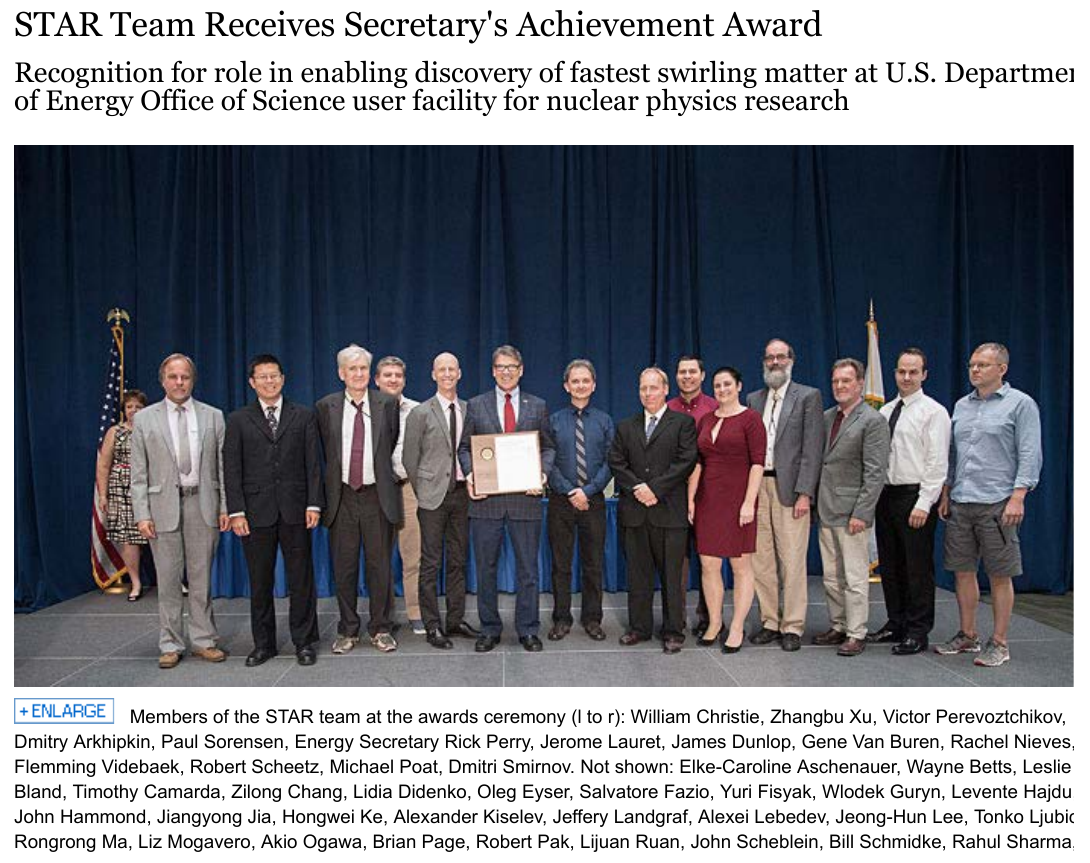}}\hspace*{0.2pc}
\end{center}\vspace*{-1.5pc}
\caption[]{\footnotesize STAR receives an award for vorticity in 2018 BUT Michael Lisa isn't there!!?}
\label{fig:STARaward}\vspace*{-0.5pc}
\end{figure}

\section{The search for the Quark Gluon Plasma at RHIC}
\begin{figure}[!h]
\begin{center}
\raisebox{0pc}{\includegraphics[width=0.70\textwidth]{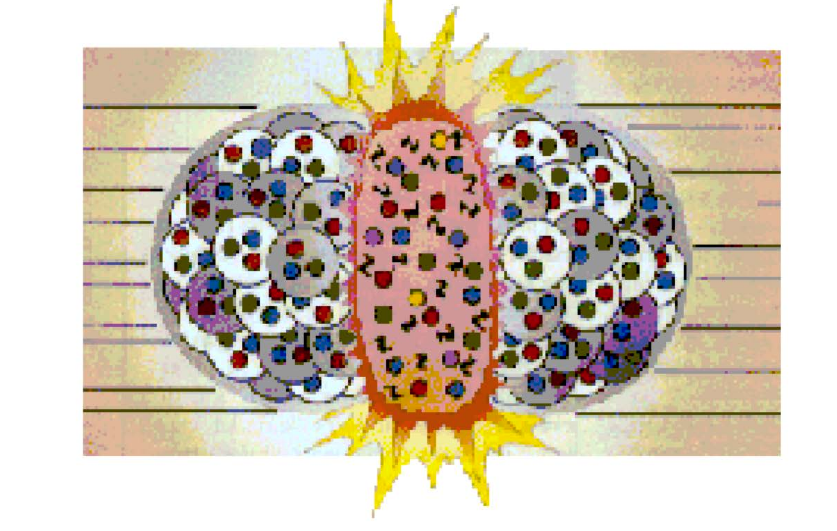}}\hspace*{0.2pc}
\end{center}\vspace*{-1.5pc}
\caption[]{\footnotesize sketch of Nucleus-Nucleus collision producing a \QGP }
\label{fig:AAcollision}\vspace*{-1.5pc}
\end{figure}

\begin{figure}[!h]
\begin{center}
\raisebox{0pc}{\includegraphics[width=0.80\textwidth]{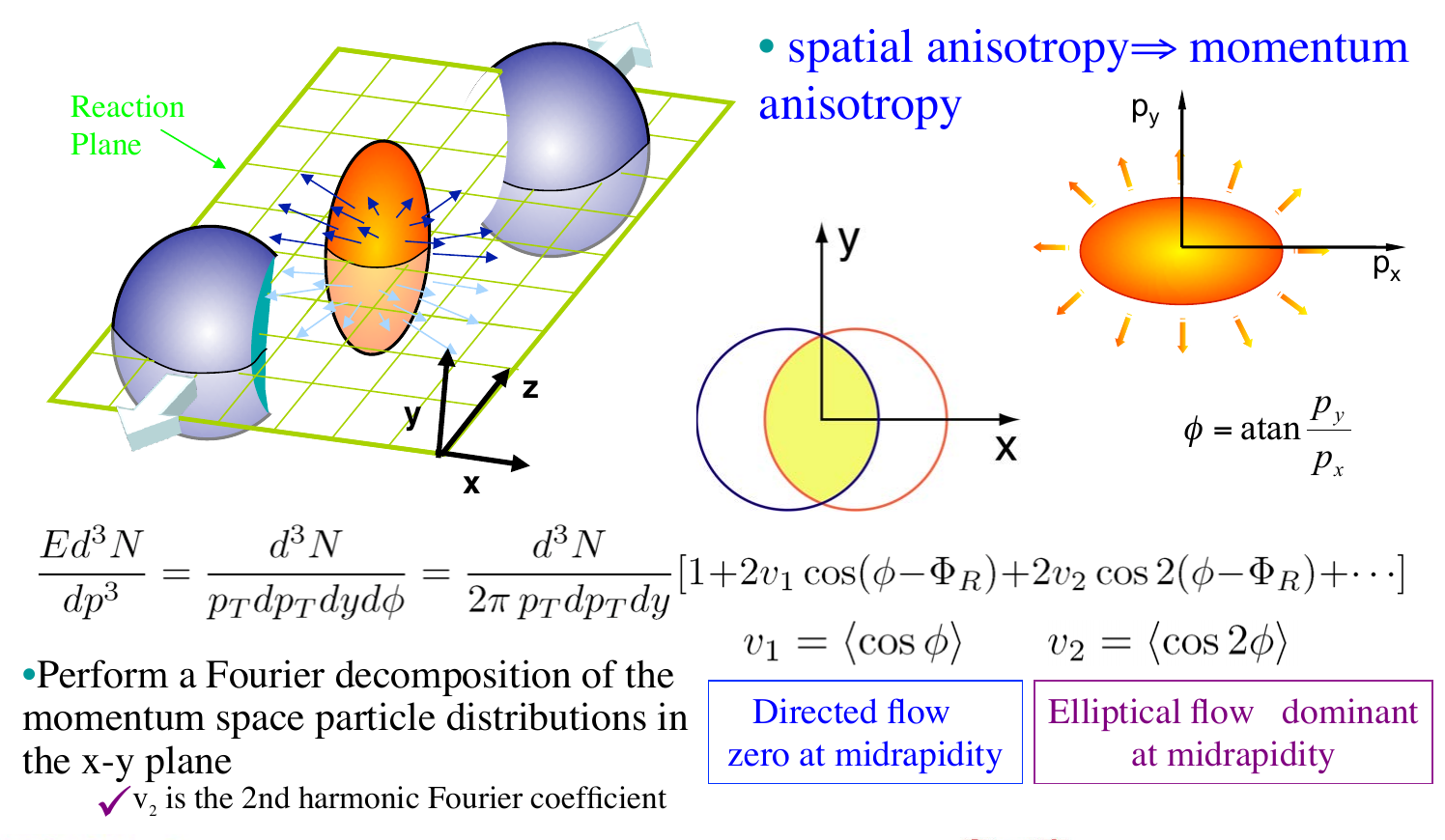}}\hspace*{0.2pc}
\end{center}\vspace*{-1.5pc}
\caption[]{\footnotesize Sketch and definitions of Elliptical flow, $v_2$}
\label{fig:flowdef}\vspace*{-0.5pc}
\end{figure}

High energy Nucleus-Nucleus collisions provide the means of creating nuclear matter in conditions of extreme temperature and density, the Quark Gluon Plasma \QGP\ (Fig.~\ref{fig:AAcollision}). At large energy or baryon density, a phase transition is expected from a state of nucleons containing confined quarks and gluons to a state of "deconfined" (from their individual nucleons) quarks and gluons covering a volume that is many units of the confinement length. 
\subsection{\normalsize Anisotropic (Elliptical) Transverse flow-an interesting complication in all A+A collisions (Fig.~\ref{fig:flowdef})}
 
\begin{figure}[!h]
\begin{center}
\raisebox{0pc}{\includegraphics[width=0.55\textwidth]{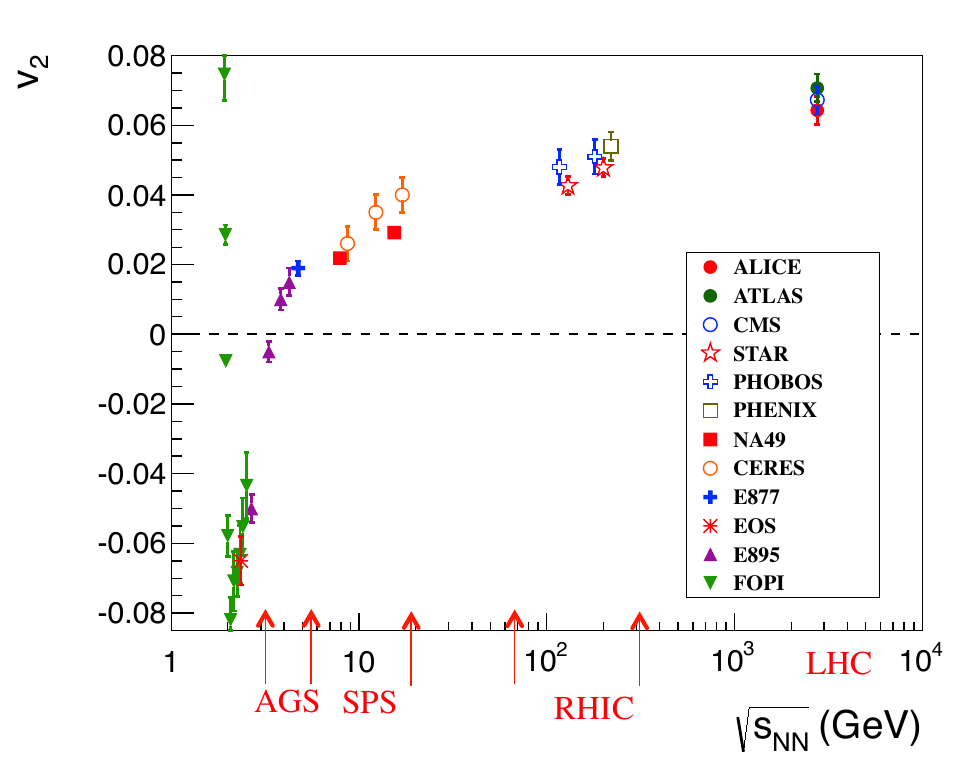}}\hspace*{0.2pc}
\end{center}\vspace*{-2.5pc}
\caption[]{\footnotesize Values of Elliptical flow ($v_2$) as a function of \sqsn from all A$+$A collision measurements. }
\label{fig:nice}\vspace*{-0.5pc}
\end{figure}
Figure \ref{fig:nice} shows that Elliptical flow ($v_2$) exists in all A$+$A collisions measured.
At very low \sqsn the main effect is from nuclei bouncing off each other and breaking to fragments. The negative $v_2$ at larger \sqsn is produced by the effective ``squeeze-out'' (in the $y$ direction) of the produced particles by slow moving minimally Lorentz-contracted spectators which block the particles emitted in the reaction plane. With increasing \sqsn, the spectators move faster and become  more contracted so the blocking stops and positive $v_2$ returns.  
\subsection{\normalsize Flow also exists in small systems and is sensitive to the initial geometry}\vspace*{-2.0pc}
\begin{figure}[!h]
\begin{center}
\raisebox{0pc}{\includegraphics[width=0.75\textwidth]{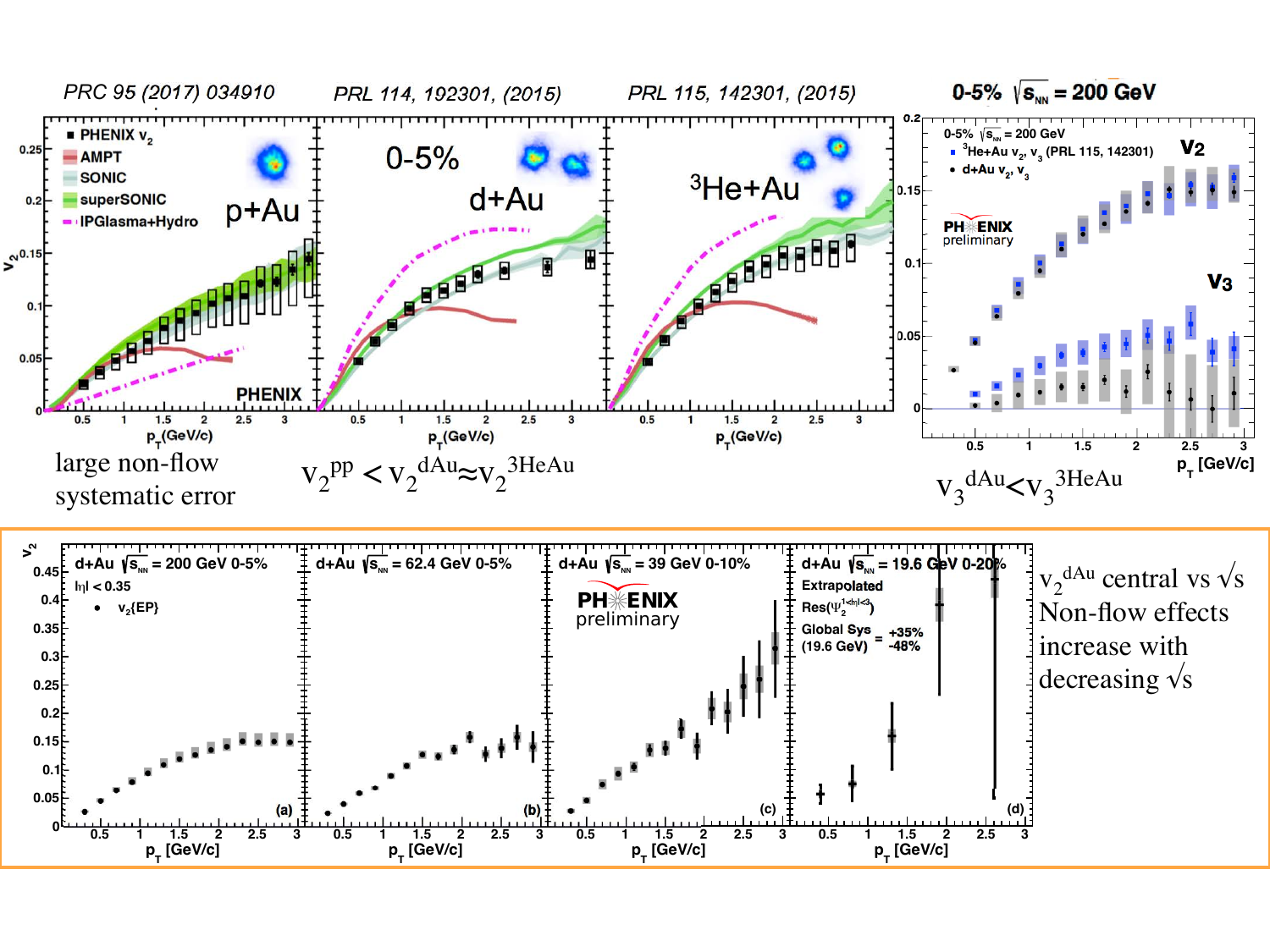}}\hspace*{0.2pc}
\end{center}\vspace*{-2.5pc}
\caption[]{\footnotesize (top) Published PHENIX $v_2$ measurements in p$+$Au, and 0-5\% central d+Au and $^3$He$+$Au collisions at \sqsn=200 GeV, with preliminary $v_2$ and $v_3$ for the d$+$Au and $^3$He$+$Au compared on the right. (bottom) PHENIX preliminary $v_2$ in d$+$Au collisions as a function of \sqsn with the centrality indicated illustrating that non-flow effects increase with decreasing \sqsn. }
\label{fig:nicer}\vspace*{-0.5pc}
\end{figure}

\begin{figure}[!h]
\begin{center}
\raisebox{0pc}{\includegraphics[width=0.78\textwidth]{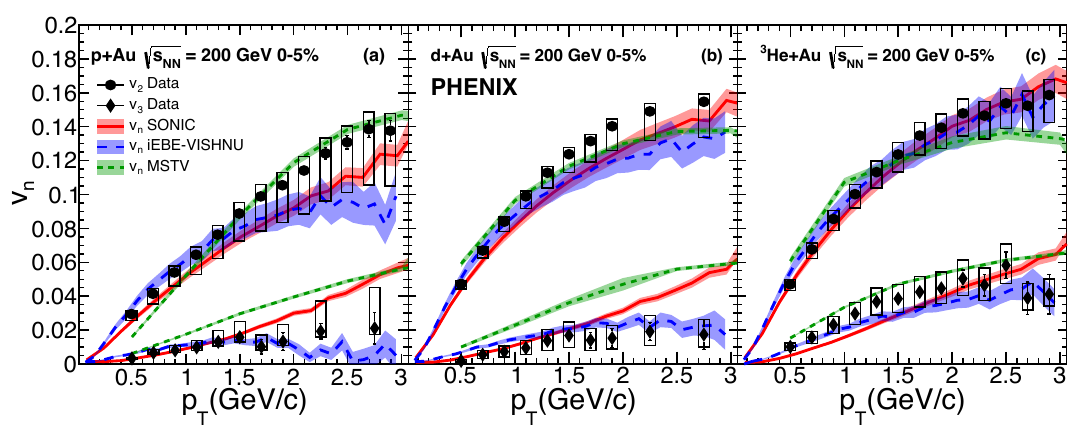}}\hspace*{0.2pc}\\
\raisebox{0pc}{\includegraphics[width=0.78\textwidth]{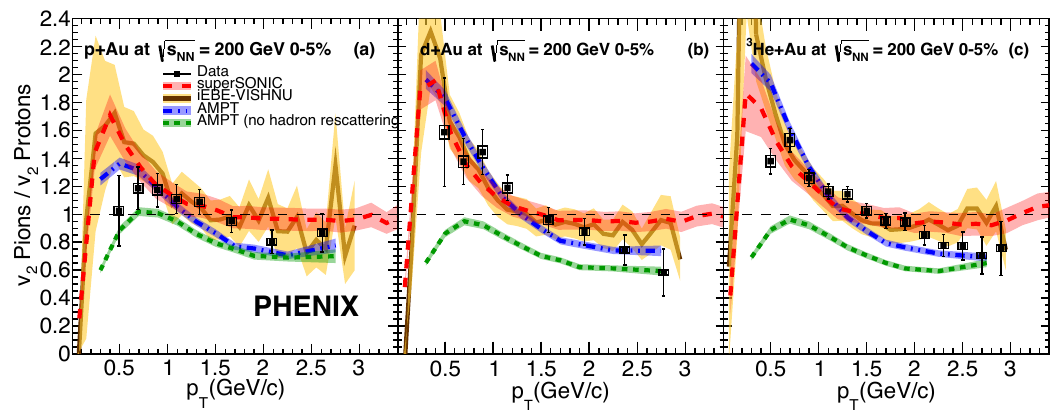}}\hspace*{0.2pc}
\end{center}\vspace*{-1.5pc}
\caption[]{\footnotesize A)(top) $v_2$ and $v_3$ in in 0-5\% central (a) p$+$Au, (b) d$+$Au, (c) ${^3}$He$+$Au collisions at \sqsn=200 GeV [\Journal{Nature Physics\ }{15} {214Ð220} {2019}]. B)(bottom) $v_2$ Pions/$v_2$ Protons in 0-5\% central (a) p$+$Au, (b) d$+$Au, (c) ${^3}$He$+$Au collisions at \sqsn=200 GeV [\Journal{PRC}{97}{064904}{2018}].  }
\label{fig:evennicer}\vspace*{-0.5pc}
\end{figure}

Fig.~\ref{fig:nicer} showed that flow exists in small p$+$Au, d$+$Au, ${^3}$He$+$Au systems with preliminary sensitivity of $v_3$ to the initial geometry. Fig.~\ref{fig:evennicer}A shows that $v_2$ is about the same in all 3 systems but $v_3$ is much larger in ${^3}$He$+$Au clearly indicating the sensitivity of flow to the initial geometry of the collision. Fig.~\ref{fig:evennicer}B shows that there is mass ordering in the flow which is strong evidence for hydrodynamics in these small systems. The solid red and dashed blue lines represent hydrodynamic predictions. These  hydrodynamical models, which include the formation of a short-lived \QGP\ droplet, provide the best simultaneous description of the measurements, strong evidence for the \QGP\ in small systems.  
\subsubsection{``It takes two to tango".--- J.~L.~Nagle {\it et al.} \Journal{PRC\ }{97}{024909}{2018}}
This is an answer to the interesting question of the minimal conditions for collectivity in small systems.
\begin{figure}[!h]
\begin{center}
\raisebox{0pc}{\includegraphics[width=0.80\textwidth]{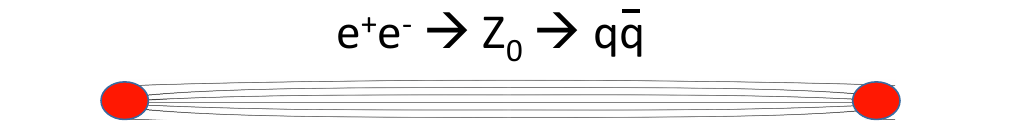}}\\\hspace*{0.2pc}
\end{center}\vspace*{-1.5pc}
\caption[]{\footnotesize  A fundamental point about \QCD\ and the string tension between the $q$ and $\bar{q}$}
\label{fig:1string}\vspace*{-0.5pc}
\end{figure}
For the case of e$^+$e$^-$ collisions in Fig.~\ref{fig:1string} utilizing the AAMPT framework and a single color string, the results indicate only a modest number of parton-parton scatterings and no observable collectivity signal.

However, a simple extension to two color strings which represent a simplified geometry in p$+$p collisions predicts finite long-range two-particle correlations (known as the ridge) and a strong $v_2$ with respect to the initial parton geometry.~\vspace*{-0.5pc}
\begin{figure}[!h]
\begin{center}
\raisebox{0pc}{\hspace*{3pc}\includegraphics[width=0.70\textwidth]{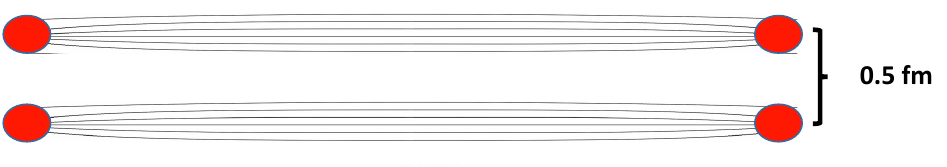}}\hspace*{0.2pc}
\end{center}\vspace*{-2.5pc}
\caption[]{\footnotesize Additional Special Case--2 Strings}\vspace*{-0.5pc}
\label{fig:2strings}\vspace*{-0.5pc}
\end{figure}
\subsubsection{A fundamental point about \QCD\ and the string tension} 
Unlike an electric or magnetic field between two sources which spreads over all space, in \QCD\ as proposed by Kogut and Susskind [\Journal{PRD\ }{9}{3501}{1974}] the color flux lines connecting two quarks or a $q-\bar{q}$
pair as in Fig.~\ref{fig:1string} are constrained in a thin tube-like region because of the three-gluon coupling. Furthermore if the field contained a constant amount
of color-field energy stored per unit length, this would provide a
linearly rising confining potential between the $q-q$ or $q-\bar{q}$ pair. 

This led to the Cornell string-like confining potential [\Journal{PRL\ }{34}{369}{1975}], which combined the
Coulomb $1/r$ dependence at short distances from vector-gluon exchange with \QCD\ coupling constant $\alpha_s(Q^2)$, and a linearly rising string-like potential, with string-tension $\sigma$, 
\begin{equation} V(r)=-{\alpha_s\over r}+\sigma r \label{eq:cornell} \end{equation}
which provided confinement at large distances (Eq.~\ref{eq:cornell}). Particles are  produced by the string breaking (fragmentation) .

\subsection{The latest discovery claims `flow' in small systems is from the \QGP\ . How did we find the \QGP\ in the first place?}
\subsubsection{$J/\psi$ Suppression, 1986}
In 1986, T. Matsui and H. Satz [\Journal{PLB\ }{178}{416}{1987}] said that due to the Debye screening of the color potential in a \QGP, charmonium production would be suppressed since the c-$\bar{\rm c}$ couldn't bind. 
With increasing temperature, $T$, in analogy to increasing $Q^2$, the strong coupling constant $\alpha_s(T)$ becomes smaller, reducing the binding energy, and the string tension, $\sigma(T)$, becomes smaller, increasing the confining radius, effectively screening the potential [\Journal{\RPP\ }{63}{1511}{2000}] 
\begin{equation} 
V(r)=-{4\over 3} {\alpha_s\over r}+\sigma r \rightarrow -{4\over 3} {\alpha_s\over r}e^{-\mu_D r}+{\sigma{(1-e^{-\mu_D r})\over \mu_D}}  \label{eq:Satz}
\end{equation}
where $\mu_D=\mu_D(T)=1/r_D$ is the Debye screening mass. 
For $r < 1/\mu_D$ a quark feels the full color charge, but for $r > 1/\mu_D$, the quark is free of the potential and the string tension, effectively deconfined. The properties of the \QGP\ can not be calculated in \QCD\ perturbation theory but only in Lattice \QCD\ Calculations [\Journal{\ARNPS\ }{65}{379}{2015}].

$J/\psi$ suppression eventually didn't work because the free $c$ and $\bar{\rm c}$ quarks recombined to make $J/\psi$'s [\Journal{PLB\ }{490}{196}{2000}]. Ask somebody from ALICE for more details.

\newpage
\subsubsection{Jet Quenching by coherent LPM radiative energy loss of a parton in the \QGP, 1997}
\begin{figure}[!h]
\begin{center}
\raisebox{0pc}{\includegraphics[width=0.70\textwidth]{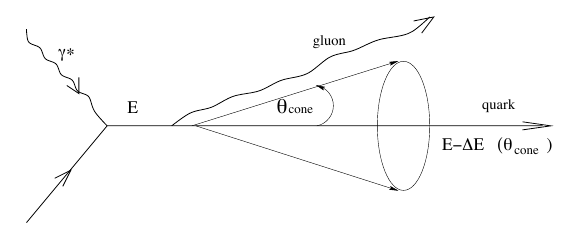}}\hspace*{0.2pc}
\end{center}\vspace*{-1.5pc}
\caption[]{\footnotesize Jet Cone of an outgoing parton with energy $E$ [BSZ arXiv:hep-ph/0002198v2]}
\label{fig:BSZcone}\vspace*{-0.5pc}
\end{figure}

In 1997, Baier, Dokshitzer, Mueller Peigne, Schiff also Zakharov (BDMPSZ), see [\Journal{\ARNPS\ }{50}{37}{2000}], said that the energy loss from coherent Landau Pomeranchuk Migdal (LPM)  radiation for hard-scattered partons exiting the \QGP\  would Òresult in an attenuation of the jet energy and a broadening of the jetsÓ. (Fig.~\ref{fig:BSZcone}).
 
As a parton from hard-scattering in the A+B collision exits through the medium it can radiate a gluon; and both continue traversing the medium. It is important to understand that ``Only the gluons radiated outside the cone defining the jet contribute to the energy loss.''  In the angular ordering of \QCD\ [\Journal{\PLB\ }{104}{161-164}{1981}], the angular cone of any further emission will be restricted to be less than that of the previous emission and will end the energy loss once inside the jet cone. This does not work in the \QGP\ so no energy loss occurs only when all gluons emitted by a parton are inside the jet cone. {\bf In addition to other issues this means that defining the jet cone is a BIG ISSUEÑ-so watch out for so-called trimming.}
\subsection{BDMPSZ--the cone, the energy loss, azimuthal broadening, is THE \QGP\ signature.}
 The energy loss of the outgoing parton, $-dE/dx$,  
per unit length ($x$) of a medium with total length $L$, is proportional to the total 4-momentum transfer-squared, $q^2(L)$, and takes the form:\vspace*{-0.5pc}
\[{-dE \over dx }\simeq \alpha_s \langle{q^2(L)}\rangle=\alpha_s\, \mu^2\, L/\lambda_{\rm mfp} 
=\alpha_s\, \hat{q}\, L\qquad \]
where $\mu$, is the mean momentum transfer per collision, and the transport coefficient $\hat{q}=\mu^2/\lambda_{\rm mfp}$ is the 4-momentum-transfer-squared to the medium per mean free path, $\lambda_{\rm mfp}$.\\[0.5pc] 

 Additionally, the accumulated momentum-squared, $\mean{p^2_{\perp W}}$ transverse to a parton traversing a length $L$ in the medium  is well approximated by
 
\[\mean{p^2_{\perp W}}\approx\langle{q^2(L)}\rangle=\hat{q}L \qquad .\]
 
\newpage

\section{Jet Quenching at RHIC, the discovery of the \QGP\  }
The energy loss of an outgoing parton with color charged fully exposed in a medium with a large density of similarly exposed color charges (i.e, a \QGP ) from Landau Pomeranchuk Migdal (LPM) coherent radiation of gluons was predicted in \QCD\ by BDMPSZ [arXiv:hep-ph/0002198v2].
\begin{figure}[!h]
\begin{center}
a)\raisebox{0pc}{\includegraphics[width=0.37\textwidth]{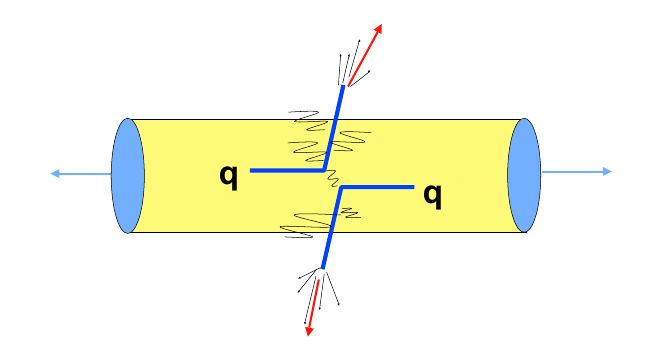}}\hspace*{0.2pc}
b)\raisebox{3pc}{\includegraphics[width=0.47\textwidth]{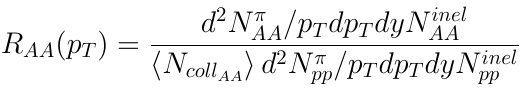}}\hspace*{0.2pc}
\end{center}\vspace*{-1.5pc}
\caption[]{\footnotesize a) Hard quark-quark scattering in an A$+$A collision with the scattered quarks passing through the medium formed in the collision. b) Nuclear modification factor $R_{AA}(p_T)$ }
\label{fig:event}\vspace*{-0.5pc}
\end{figure}

Hard scattered partons (Fig.~\ref{fig:event}a) lose energy going through the medium so that there are fewer partons or jet fragments at a given $p_T$. The ratio of the measured semi-inclusive yield of, for example, pions in a given A$+$A centrality class divided by the semi-inclusive yield in a p$+$p collision times the number of A$+$A collisions $\mean\Ncoll$ in the centrality-class is given by the nuclear modification factor, $R_{AA}$ (Fig.~\ref{fig:event}b), which equals 1 for no energy loss. 

\begin{figure}[!h]
\begin{center}
\raisebox{0pc}{\includegraphics[width=0.85\textwidth]{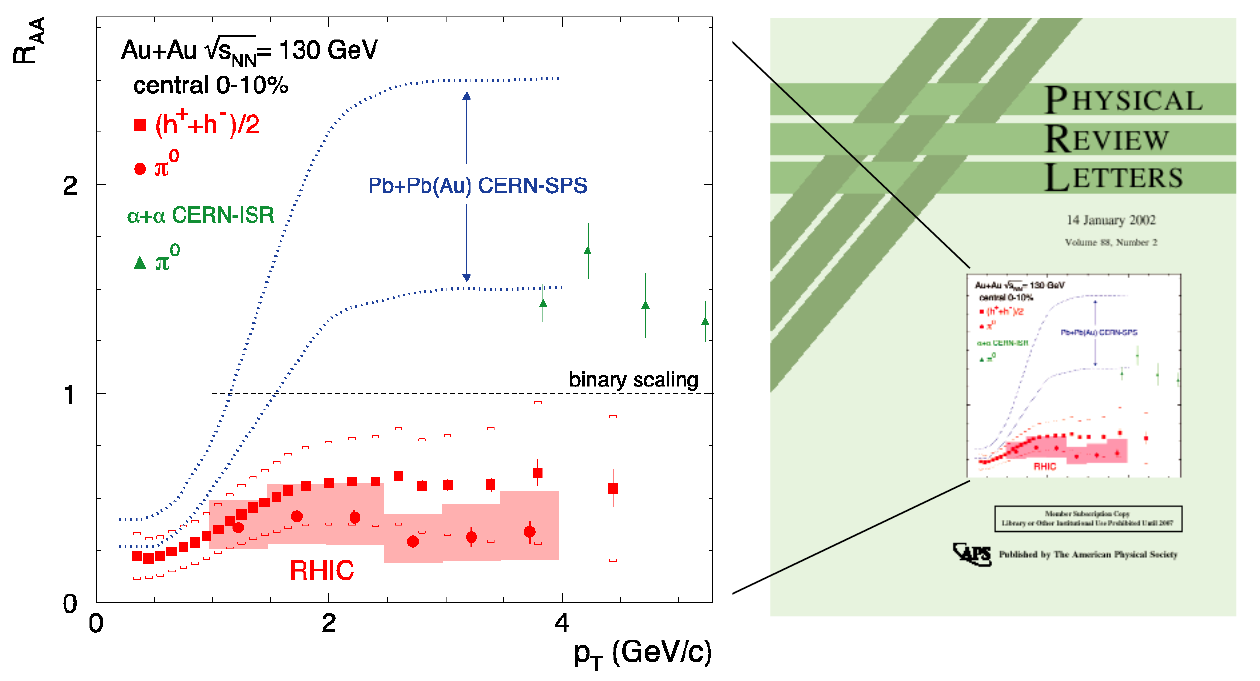}}\hspace*{0.2pc}\\
\end{center}\vspace*{-1.5pc}
\caption[]{\footnotesize (left) Hadron suppression $R_{AA}$ in Au$+$Au at \sqsn=130 GeV by PHENIX at RHIC compared to enhancement at \sqsn=17 GeV in Pb$+$Pb at the CERN SpS. (right) Plot is from the cover of PRL.   }
\label{fig:QGPdiscovery}\vspace*{-0.5pc}
\end{figure}

PHENIX discovered Jet Quenching of hadrons at RHIC in 2001 [\Journal{PRL}{88}{022301}{2002}]
(Fig.~\ref{fig:QGPdiscovery}). Pions at large $p_T>2$ GeV/c are suppressed in Au$+$Au at \sqsn=130 GeV compared to the enhancement found at the CERN SpS at \sqsn=17 GeV. This is the first regular publication from a RHIC experiment to reach 1000 citations.
\subsection{Status of $R_{AA}$ in Au+Au at \sqsn=200 GeV}
\begin{figure}[!h]
\begin{center}\vspace*{-1.5pc}
\raisebox{0pc}{\includegraphics[width=0.80\textwidth]{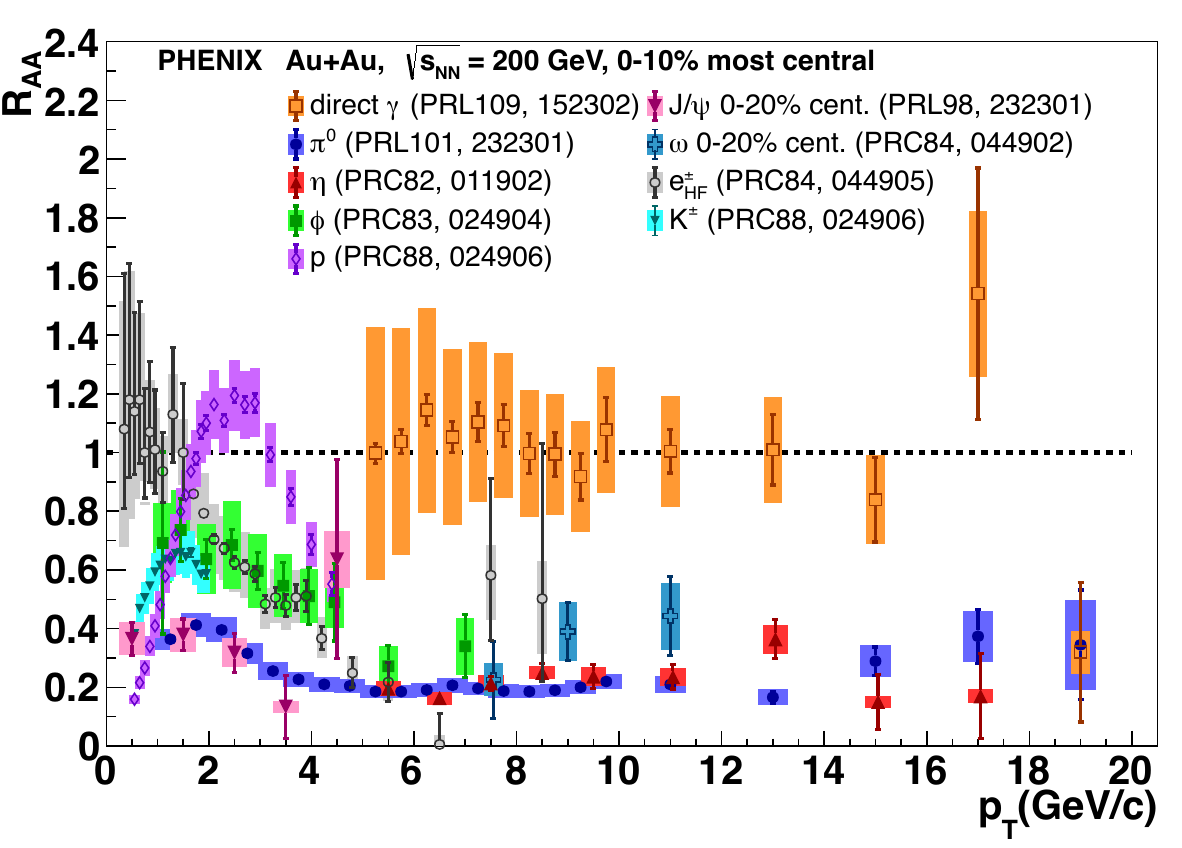}}\hspace*{0.2pc}
\end{center}\vspace*{-1.5pc}
\caption[]{\footnotesize Published PHENIX measurements of $R_{AA}$ with references.}
\label{fig:PXRAA}\vspace*{-0.5pc}
\end{figure}
Figure \ref{fig:PXRAA} shows the suppression of all identified hadrons, as well as $e^{\pm}$ from $c$ and $b$ quark decay, with $p_T>2$ GeV/c measured by PHENIX until 2013. One exception is  the enhancement of protons for $2<p_T<4$ GeV/c which are then suppressed at larger $p_T$. Particle Identification is crucial for these measurements since all particles behave differently. The only particle that shows no-suppression is the direct single $\gamma$ (from the \QCD\ reaction $g+q\rightarrow \gamma +q$) which shows that the medium produced at RHIC is the strongly interacting \QGP\ since $\gamma$ rays only interact electromagnetically. 

\subsection{Recent measurements to test the second BDMPSZ prediction.}
\begin{figure}[!h]
\begin{center}
\raisebox{0pc}{\includegraphics[width=0.70\textwidth]{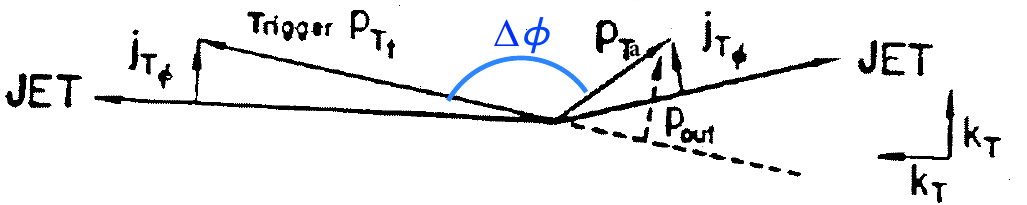}}\hspace*{0.2pc}
\end{center}
\caption[]{\footnotesize Sketch of a di-jet looking down the beam axis. The $k_T$ from the two jets add randomly and are shown with one $k_T$ perpendicular to the scattering plane which makes the jets acoplanar in azimuth and the other $k_T$ parallel to the trigger jet which makes the jets unequal in energy. Also $x_E=p_{Ta}\cos(\pi-\Delta\phi)/p_{Tt}$. The formula for calculating $k_T$ from di-hadron correlations is given in [\Journal{PRD\ }{74}{072002}{2006}].}
\label{fig:kTdiagram}
\end{figure}

\noindent{\color{PineGreen}(1)} The energy loss of the outgoing parton, $-dE/dx$,  
per unit length ($x$) of a medium with total length $L$, is proportional to the total 4-momentum transfer-squared, $q^2(L)$, and takes the form:\vspace*{-0.5pc}
\[{-dE \over dx }\simeq \alpha_s \langle{q^2(L)}\rangle=\alpha_s\, \mu^2\, L/\lambda_{\rm mfp} 
=\alpha_s\, \hat{q}\, L\qquad \]
where $\mu$, is the mean momentum transfer per collision, and the transport coefficient $\hat{q}=\mu^2/\lambda_{\rm mfp}$ is the 4-momentum-transfer-squared to the medium per mean free path, $\lambda_{\rm mfp}$.\\[0.5pc] 
\noindent{\color{PineGreen}(2)} Additionally, the accumulated momentum-squared, $\mean{p^2_{\perp W}}$ transverse to a parton traversing a length $L$ in the medium  is well approximated by\vspace*{-0.5pc} 
\begin{equation}
\mean{p^2_{\perp W}}\approx\langle{q^2(L)}\rangle=\hat{q}\, L \qquad  \mean{\hat{q} L}=\mean{k_{T}^2}_{AA}-\mean{k{'}_{T}^2}_{pp} \label{Eq:qhatL} \end{equation}. 
Although only the component of $\mean{p^2_{\perp W}}$ $\perp$ to the scattering plane affects $k_T$ (Fig.~\ref{fig:kTdiagram}) the azimuthal broadening of the di-jet is caused by the random sum of the azimuthal components $\mean{p^2_{\perp W}}/2$ from each outgoing di-jet or $\mean{p^2_{\perp W}}=\hat{q}\, L$. 

From the values of $R_{AA}$ observed at RHIC (after 12 years) the JET Collaboration [\Journal{\PRC\ }{90}{014909}{2014}] has found that $\hat{q}=1.2\pm 0.3$ GeV$^2$/fm at RHIC, $1.9\pm 0.6$ at LHC at an initial time $\tau_0=0.6$ fm/c; but nobody has yet measured the azimuthal broadening predicted. Before proceeding, one has to know the meaning of $k_T$ defined by Feynman, Field and Fox in [\Journal{NPB\ }{129}{1}{1977}] as the transverse momentum of a parton in a nucleon (Fig.~\ref{fig:kTdiagram}). 
\subsubsection{The key new idea of $\mean{k{'}_{T}^2}_{pp}$ instead of $\mean{k_{T}^2}_{pp}$ in Eq.~\ref{Eq:qhatL}}
The di-hadron correlations of $p_{Ta}$ with $p_{Tt}$ (Fig.~\ref{fig:kTdiagram}) are measured in p$+$p and Au$+$Au collisions. The parent jets in the original Au$+$Au collision as measured in p$+$p will both lose energy passing through the medium but the azimuthal angle between the jets should not change unless the medium induces multiple scattering from $\hat{q}$. Thus the calculation of $k{'}_{T}$ from the dihadron p+p mesurement to compare with Au+Au measurements with the same di-hadron $p_{Tt}$ and $p_{Ta}$ must use the value of $\hat{x}_h$ and $\mean{z_t}$ of the parent jets in the A+A collision. 
The variables are $x_h\equiv p_{Ta}/p_{Tt},  \hat{x}_h\equiv \hat{p}_{Ta}/\hat{p}_{Tt},  \mean{z_t}\equiv p_{Tt}/\hat{p}_{Tt}$ where e.g. $p_{Tt}$ is the trigger particle transverse momentum and $\hat{p}_{Tt}$ means the trigger jet transverse momentum. \vspace*{0.5pc}

The same values of $\hat{x}_h$, and $\mean{z_t}$ in Au$+$Au and p$+$p give the cool result   [\Journal{PLB\ }{771}{553}{2017}]:
\begin{equation} \mean{\hat{q} L}=\left[\frac{\hat{x}_h}{\mean{z_t}}\right]^2 \;\left[\frac{\mean{p^2_{\rm out}}_{AA} - \mean{p^2_{\rm out}}_{pp}}{x_h^2}\right] 
\label{eq:qhatLparticle} \end{equation}

 For di-jet measurements, the formula is  even simpler:\vspace*{0.5pc} \\ i) $x_h \equiv \hat{x}_h$ because the trigger and away `particles' are the jets; ii) $\mean{z_t}\equiv 1$ because the trigger `particle' is the entire jet not a fragment of the  jet;\\ iii) $\mean{p^2_{\rm out}}=\hat{p}_{Ta}^2 \sin^2(\pi-\Delta\phi)$. This reduces the formula for di-jets to: \vspace*{-0.5pc}
\begin{equation}\mean{\hat{q} L}= \left[{\mean{p^2_{\rm out}}_{AA} - \mean{p^2_{\rm out}}_{pp}}\right]
=\hat{p}_{Ta}^2 \left[ {\mean{\sin^2(\pi-\Delta\phi)}_{AA} - \mean{\sin^2(\pi-\Delta\phi)}_{pp}}\right]\label{eq:qhatLjet}\end{equation}
\pagebreak

\subsubsection{A test of Eq.~\ref{eq:qhatLjet} for $\mean{\hat{q} L}$}
\begin{figure}[!h]
\begin{center}
\raisebox{0pc}{\includegraphics[width=0.66\textwidth]{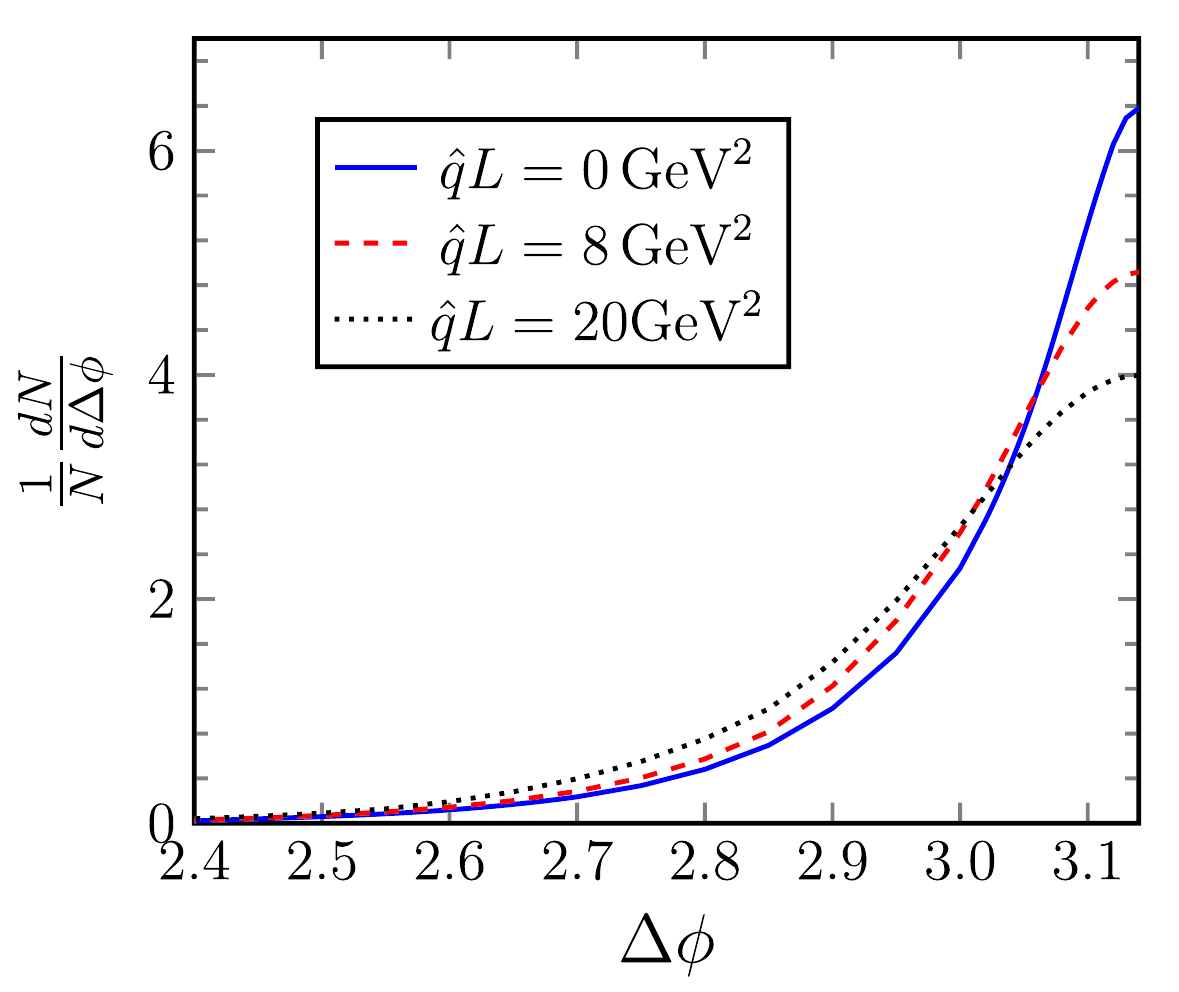}}
\end{center}\vspace*{-1.5pc}
\caption[]{\footnotesize Prediction of folded away azimuthal width of 35 GeV/c Jets at RHIC for several values of $\hat{q} L$}
\label{fig:Muellerclean}\vspace*{-0.5pc}
\end{figure}

Al Mueller et al. [\Journal{PLB\ }{763}{208}{2016}] gave a prediction for the azimuthal broadening of dijet angular correlations for 35 GeV jets at RHIC (Fig.~\ref{fig:Muellerclean}).
To check my Eq.~\ref{eq:qhatLjet}, I measured the half width at half maximum (HWHM), which equals $1.175\sigma$ for a Gaussian, for each curve in Fig.~\ref{fig:Muellerclean},  and calculated  $(\sigma \times 35)^2$  to get $\mean{p^2_{\rm out}}$ for each $\hat{q}L$, and used Eq.~\ref{eq:qhatLjet} to get 9.6 GeV$^2$ and 21.5 GeV$^2$ respectively for the 8 GeV$^2$ and 20 GeV$^2$ plots. This is an excellent result considering that I had to measure the HWHMs from Fig.~\ref{fig:Muellerclean} with a pencil and ruler. 
\subsubsection{How to calculate $\hat{q}L$ with Eq.~\ref{eq:qhatLparticle} from di-hadron measurements }
The determination of the required quantities is well known to older PHENIXians who have read [\Journal{PRD\ }{74}{072002}{2006}] or my book [Rak \& Tannenbaum, High pT physics in the Heavy Ion Era-Cambridge 2013] as outlined below:

(A) $\mean{z_t}$ is calculated from the Bjorken parent-child relation and `trigger bias' [\Journal{Phys. Rep.\ }{48}{285}{1978}], also see \Journal{PRD\ }{81}{012002}{2010};

(B) The energy loss of the trigger jet from p$+$p to Au$+$Au can be measured by the shift in the $p_T$ spectra [\Journal{PRC\ }{87}{034911}{2013}]; 

(C) $\hat{x}_h$, the ratio of the away-jet to the trigger jet transverse momenta can be measured by the away particle $p_{Ta}$ distribution for a given trigger particle $p_{Tt}$ taking  $x_E=x_h\cos{\Delta\phi}\approx x_h=p_{Ta}/p_{Tt}$:

    \begin{equation}
\left.{dP_{\pi} \over dx_E}\right|_{p_{T_t}}  = {N\,(n-1)}{1\over\hat{x}_h} {1\over {(1+ {x_E \over{\hat{x}_h}})^{n}}} \qquad . 
\qquad  
\label{eq:condxe2}
\end{equation}

\subsubsection{Example: $\hat{x}_h$ from fits to the PHENIX data from [\Journal{PRL\ }{104}{252301}{2010}}
\begin{figure}[!h]\vspace*{-1.0pc}
\begin{center}
\raisebox{0pc}{\includegraphics[width=0.40\textwidth]{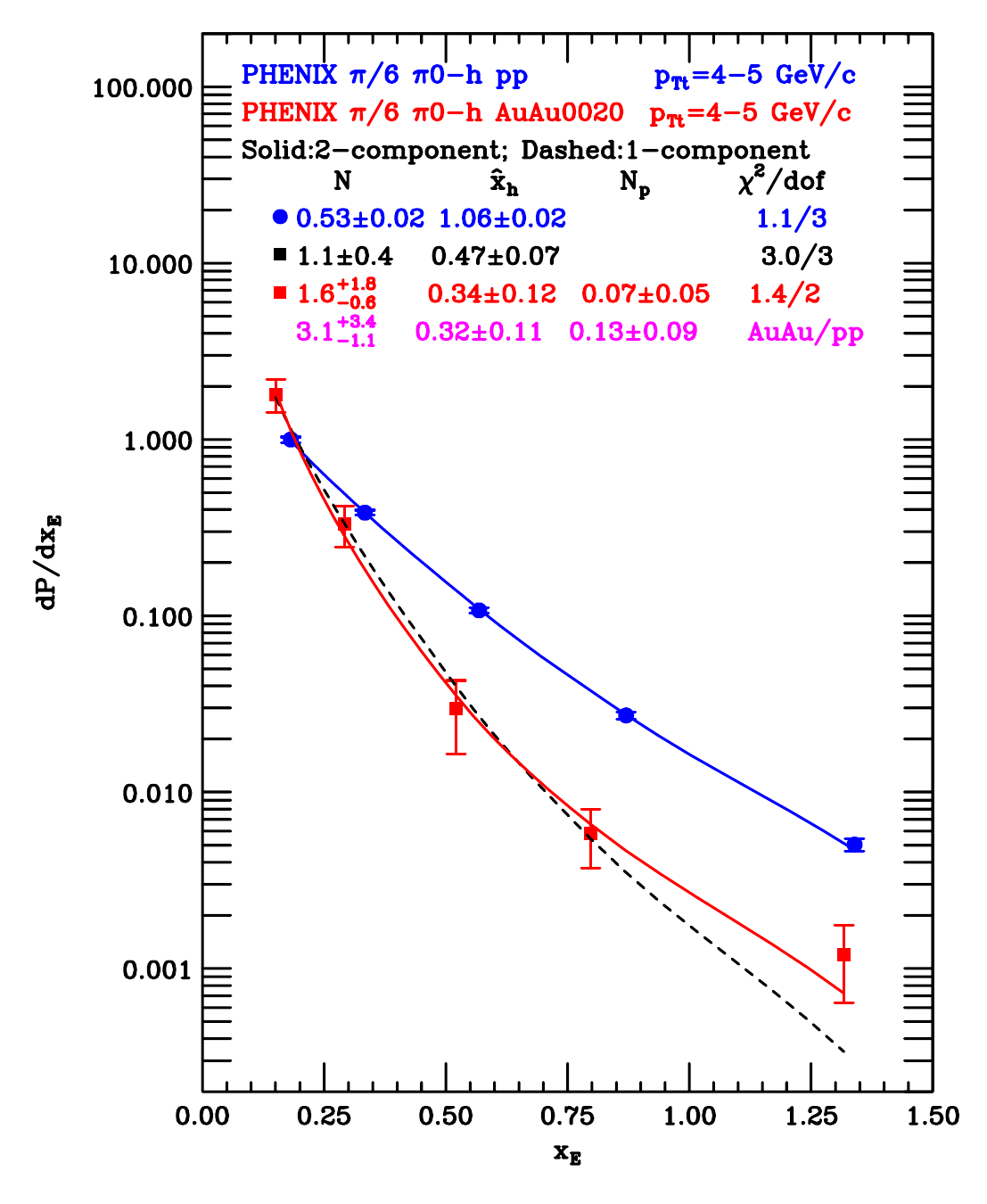}}\hspace*{0.2pc}
\raisebox{0pc}{\includegraphics[width=0.40\textwidth]{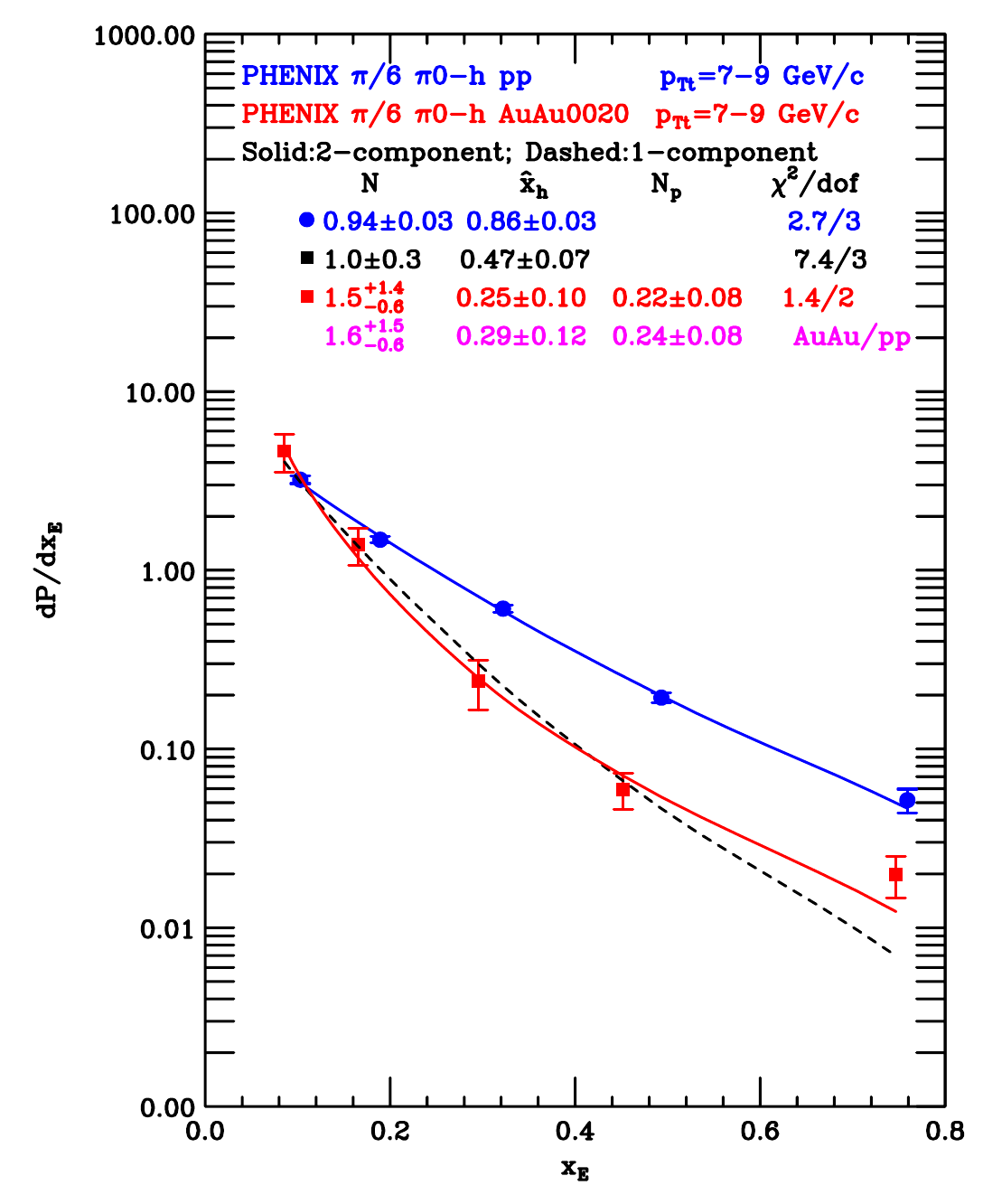}}\hspace*{0.2pc}
\end{center}\vspace*{-2.0pc}
\caption[]{\footnotesize Fit to $x_E$ distributions for $\pi^0-h$ correlation in p$+$p and Au$+$Au 0-20\% central collisions using Eq.~\ref{eq:condxe2} with the results indicated: (left) $4<p_{Tt}<5$ GeV/c; (rght) $7<p_{Tt}<9$ GeV/c;}
\label{fig:xTdist}\vspace*{-0.5pc}
\end{figure}
The fits in Fig.~\ref{fig:xTdist} work very well, with excellent $\chi^2$/dof. However it is important to notice that the dashed curve in Au$+$Au doesn't fit the data as well as the solid red curve which is the sum of Eq.~\ref{eq:condxe2} with free parameters + a second term with the form of Eq.~\ref{eq:condxe2} but with the $\hat{x}_h$ fixed at the p$+$p value. It is also important to note that the solid red curve between the highest Au$+$Au data points is notably parallel to the p$+$p curve. A possible explanation is that in this region, which is at a fraction $\approx 1$\% of the $dP/dx_E$ distribution, the highest $p_{Ta}$ fragments are from jets that don't lose energy in the \QGP\ .\vspace*{-1.0pc}
\subsubsection{Results from STAR $\pi^0-h$ and $\gamma-h$ correlations [\Journal{PLB\ }{760}{689}{2016}]}
    \begin{table}[!h]\vspace*{-2.0pc}
\begin{center}\footnotesize
\caption[]{\footnotesize $\hat{q}L$ result table for STAR $\pi^0$-h: $12<p_{Tt}<20$ GeV/c 00-12\% Centrality}
{\begin{tabular}{ccccccc} 
\hline
STAR PLB760\\
\hline
$\sqsn=200$ &$\mean{p_{Tt}}$ & $\mean{p_{Ta}}$ & $\mean{z_t}$ & $\hat{x}_h$&$\mean{p^2_{\rm out}}$ &$\sqrt{\mean{k^2_T}}$\\ 
\hline
Reaction &   GeV/c &  GeV/c&     & GeV/c &   & GeV/c\\ 
\hline
p$+$p&14.71&1.72&$0.80\pm 0.05$&$0.84\pm 0.04$&$0.263\pm0.113$&$2.34\pm 0.34$\\
\hline
p$+$p&14.71&3.75&$0.80\pm 0.05$&$0.84\pm 0.04$&$0.576\pm0.167$&$2.51\pm 0.31$\\
\hline
Au$+$Au 00-12\%&14.71&1.72&$0.80\pm0.05$&$0.36\pm 0.05$&$0.547\pm 0.163$ &$2.28\pm 0.35$\\
\hline
Au$+$Au 00-12\%&14.71&3.75&$0.80\pm0.05$&$0.36\pm 0.05$&$0.851\pm0.203$&$1.42\pm 0.22$\\
\hline
p$+$p comp&14.71&1.72&$0.80\pm0.05$&$0.36\pm 0.05$&$0.263\pm0.113$ &$1.006\pm 0.18$\\
\hline
p$+$p comp&14.71&3.75&$0.80\pm0.05$&$0.36\pm 0.05$&$0.576\pm0.167$&$1.076\pm 0.18$\\
\hline\\[-1.0pc]
\hline
 &   &  &     & &   & $\mean{\hat{q} L}$ GeV$^2$ \\
 \hline
Au$+$Au 00-12\%&14.71&1.72& &$$&&$4.21\pm3.24 $*\\
\hline
Au$+$Au 00-12\%&14.71&3.75&  & $$&&$0.86\pm0.87$*\\
\hline\\[-1.0pc] 
\hline
\end{tabular}} \label{tab:STARPLB771}
\end{center}\vspace*{-0.5pc}
\end{table}
\normalsize\vspace*{-1.0pc}
Table~\ref{tab:STARPLB771} is a table of results of my published calculation [\Journal{PLB\ }{771}{553}{2017}] of $\mean{\hat{q}L}$ from the STAR data. The errors on the STAR $\mean{\hat{q}L}$ here (with the *) are much larger than stated in my published calculation because I made a trivial mistake which is corrected here. Also the new values of $\mean{\hat{q}L}$ reflect that Eq.~\ref{eq:qhatLparticle} defines $\mean{\hat{q}L}$ not $\mean{\hat{q}L}/2.$

\subsection{Some $\mean{\hat{q}L}$ results from PHENIX [\Journal{PRL\ }{104}{252301}{2010}]}
\begin{figure}[!h]
\begin{center}\vspace*{-1.0pc}
\raisebox{0pc}{\includegraphics[width=0.65\textwidth]{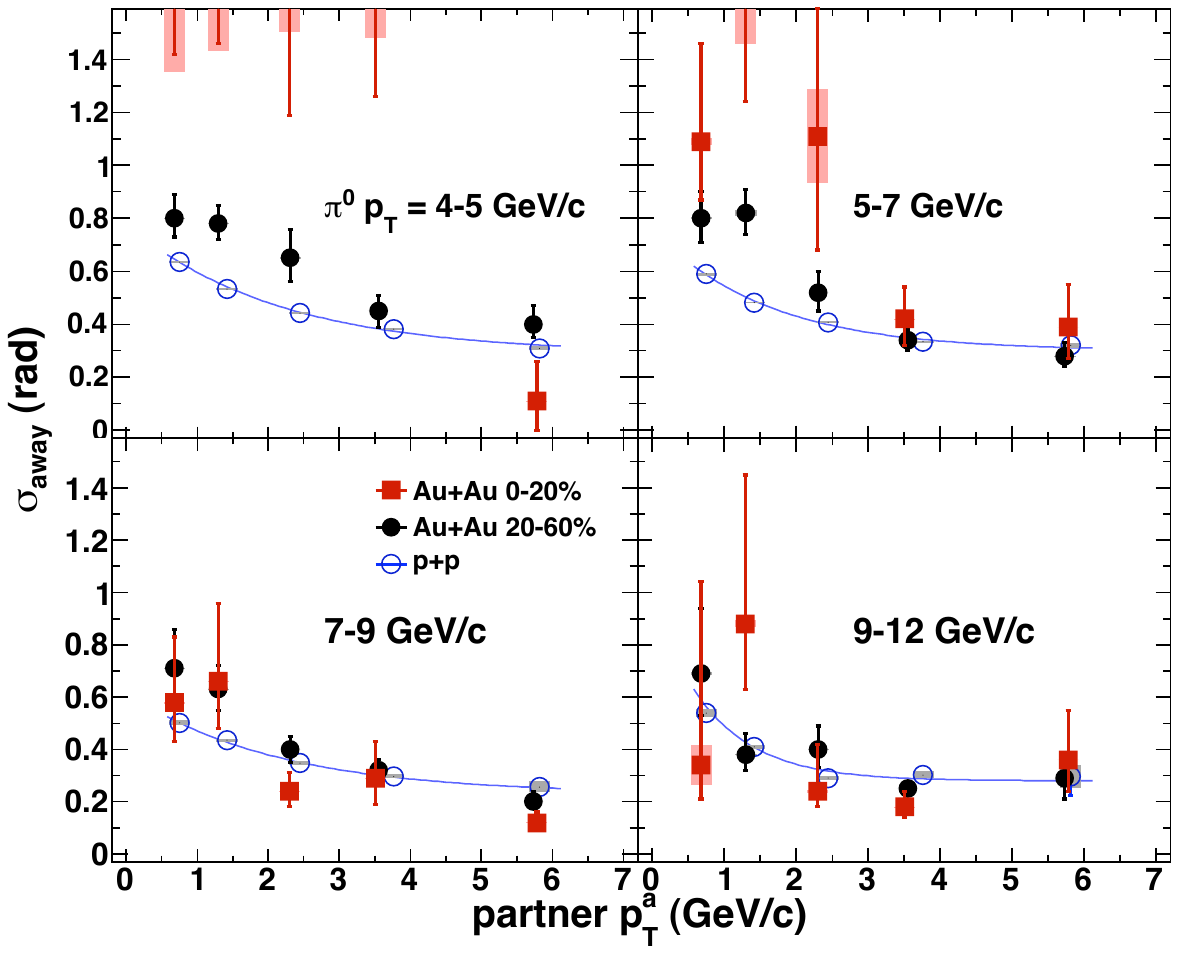}}\hspace*{0.2pc}
\end{center}\vspace*{-1.5pc}
\caption[]{\footnotesize Away widths from $\pi^0-h$ correlations as function of partner $p_T$, i.e. $p_{Ta}$, in Au$+$Au 0-20\% and 20-60\% and p$+$p collisions at \sqsn=200 GeV for 4 ranges of trigger $p_{Tt}$ indicated.}
\label{fig:ppg106sigma}\vspace*{-0.5pc}
\end{figure}
The away widths from PHENIX $\pi^0-h$ correlations [\Journal{PRL\ }{104}{252301}{2010}] are shown in Fig.~\ref{fig:ppg106sigma} with the calculated $\hat{q}L$ values for $\pi^0-h$ GeV/c 20-60\% centrality $5<p_{Tt}<7$ GeV/c shown in Table ~\ref{tab:PX-PRL104-5-7-20-60} and $7<p_{Tt}<9$ GeV/c in Table ~\ref{tab:PX-PRL104-7-9-20-60}.
     \begin{table}[!h]\vspace*{-1.0pc}
\begin{center}\footnotesize
\caption[]{\footnotesize$\hat{q}L$ result table for PHENIX $\pi^0$-h: $5<p_{Tt}<7$ GeV/c 20-60\% Centrality}
{\begin{tabular}{ccccccc} 
\hline
PHENIX PRL104\\
\hline
$\sqsn=200$ &$\mean{p_{Tt}}$ & $\mean{p_{Ta}}$ & $\mean{z_t}$ & $\hat{x}_h$&$\mean{p^2_{\rm out}}$ &$\sqrt{\mean{k^2_T}}$\\ 
\hline
Reaction &   GeV/c &  GeV/c&     & GeV/c &   & GeV/c\\ 
\hline
p$+$p&5.78&1.42&$0.60\pm 0.06$&$0.96\pm 0.02$&$0.434\pm0.010$&$3.13\pm 0.37$\\%
\hline
p$+$p&5.78&2.44&$0.60\pm 0.06$&$0.96\pm 0.02$&$0.934\pm0.031$&$3.18\pm 0.34$\\%
\hline
p$+$p&5.78&3.76&$0.60\pm 0.06$&$0.96\pm 0.02$&$1.523\pm0.061$&$2.74\pm 0.29$\\%
\hline
p$+$p&5.78&5.82&$0.60\pm 0.06$&$0.96\pm 0.02$&$3.339\pm0.351$&$2.73\pm 0.32$\\%
\hline 
Au$+$Au 20-60\%&5.78&1.30&$0.62\pm 0.06$&$0.69\pm 0.05$&$0.867\pm 0.116$ &$4.04\pm 0.61$\\%
\hline
Au$+$Au 20-60\%&5.78&2.31&$0.62\pm 0.06$&$0.69\pm 0.05$&$1.291\pm0.308$&$2.88\pm 0.54$\\%
\hline
Au$+$Au 20-60\%&5.78&3.55&$0.62\pm 0.06$&$0.69\pm 0.05$&$1.370\pm0.249$&$1.90\pm 0.32$\\%
\hline
Au$+$Au 20-60\%&5.78&5.73&$0.62\pm 0.06$&$0.69\pm 0.05$&$2.562\pm0.620$&$1.66\pm 0.31$\\%
\hline
p$+$p comp&5.78&1.30&$0.62\pm 0.06$&$0.69\pm 0.05$&$0.434\pm 0.010$ &$2.39\pm 0.32$\\%
\hline
p$+$p comp&5.78&2.31&$0.62\pm 0.06$&$0.69\pm 0.05$&$0.934\pm0.031$&$2.34\pm 0.29$\\%
\hline
p$+$p comp&5.78&3.55&$0.62\pm 0.06$&$0.69\pm 0.05$&$1.522\pm0.061$&$2.03\pm 0.25$\\%
\hline
p$+$p comp&5.783&5.73&$0.62\pm 0.06$&$0.69\pm 0.05$&$3.339\pm0.351$&$1.93\pm 0.26$\\%
\hline\\[-1.0pc]
\hline
 &   &  &     & & $\mean{\hat{q} L}.01$& $\mean{\hat{q} L}$ GeV$^2$ \\
 \hline
Au$+$Au 20-60\%&5.78&1.30& &$$&$6.9\pm3.6 $&$10.6\pm3.8 $\\
\hline
Au$+$Au 20-60\%&5.78&2.31&  & $$&$2.3\pm 2.1$&$2.8\pm 2.4$\\
\hline
Au$+$Au 20-60\%&5.78&3.55&  & $$&$0.35\pm 0.93$&$-0.5\pm 0.9$\\
\hline
Au$+$Au 20-60\%&5.78&5.73& & $$&$-0.75\pm 1.0$&$-1.0\pm 0.9$\\
\hline\\[-1.0pc] 
\hline
\end{tabular}} \label{tab:PX-PRL104-5-7-20-60}
\end{center}\vspace*{-0.5pc}
\end{table}\normalsize

     \begin{table}[!t]\vspace*{-1.0pc}
\begin{center}\footnotesize
\caption[]{\footnotesize$\hat{q}L$ result table for PHENIX $\pi^0$-h: $7<p_{Tt}<9$ GeV/c 20-60\% Centrality}
{\begin{tabular}{ccccccc} 
\hline
PHENIX PRL104\\
\hline
$\sqsn=200$ &$\mean{p_{Tt}}$ & $\mean{p_{Ta}}$ & $\mean{z_t}$ & $\hat{x}_h$&$\mean{p^2_{\rm out}}$ &$\sqrt{\mean{k^2_T}}$\\ 
\hline
Reaction &   GeV/c &  GeV/c&     & GeV/c &   & GeV/c\\ 
\hline
p$+$p&7.83&1.42&$0.64\pm 0.06$&$0.86\pm 0.03$&$0.360\pm0.017$&$2.98\pm 0.41$\\%
\hline
p$+$p&7.83&2.44&$0.64\pm 0.06$&$0.86\pm 0.03$&$0.694\pm0.048$&$2.99\pm 0.34$\\%
\hline
p$+$p&7.83&3.76&$0.64\pm 0.06$&$0.86\pm 0.03$&$1.213\pm0.109$&$2.76\pm 0.32$\\%
\hline
p$+$p&7.83&5.82&$0.64\pm 0.06$&$0.86\pm 0.03$&$2.177\pm0.424$&$2.48\pm 0.38$\\%
\hline 
Au$+$Au 20-60\%&7.83&1.30&$0.66\pm 0.06$&$0.62\pm 0.04$&$0.548\pm 0.107$ &$3.35\pm 0.64$\\%
\hline
Au$+$Au 20-60\%&7.83&2.31&$0.66\pm 0.06$&$0.62\pm 0.04$&$0.803\pm0.177$&$2.45\pm 0.46$\\%
\hline
Au$+$Au 20-60\%&7.83&3.55&$0.66\pm 0.06$&$0.62\pm 0.04$&$1.237\pm0.232$&$2.08\pm 0.34$\\%
\hline
Au$+$Au 20-60\%&7.83&5.73&$0.66\pm 0.06$&$0.62\pm 0.04$&$1.300\pm0.350$&$1.29\pm 0.27$\\%
\hline
p$+$p comp&7.83&1.30&$0.66\pm 0.06$&$0.62\pm 0.04$&$0.360\pm 0.017$ &$2.28\pm 0.33$\\%
\hline
p$+$p comp&7.83&2.31&$0.66\pm 0.06$&$0.62\pm 0.04$&$0.694\pm0.048$&$2.22\pm 0.28$\\%
\hline
p$+$p comp&7.83&3.55&$0.66\pm 0.06$&$0.62\pm 0.04$&$1.213\pm0.109$&$2.05\pm 0.26$\\%
\hline
p$+$p comp&7.83&5.73&$0.66\pm 0.06$&$0.62\pm 0.04$&$2.177\pm0.424$&$1.76\pm 0.28$\\%
\hline\\[-1.0pc]
\hline
 &   &  &     & & $\mean{\hat{q} L}.01$& $\mean{\hat{q} L}$ GeV$^2$ \\
 \hline
Au$+$Au 20-60\%&7.83&1.30& &$$&$9.3\pm 6.3 $&$6.0\pm3.7$\\
\hline
Au$+$Au 20-60\%&7.83&2.31&  & $$&$2.4\pm 2.2$&$1.1\pm 1.9$\\
\hline
Au$+$Au 20-60\%&7.83&3.55&  & $$&$1.0\pm 1.2$&$0.11\pm 1.1$\\
\hline
Au$+$Au 20-60\%&7.83&5.73& & $$&$-1.2\pm1.0 $&$-1.4\pm 1.0$\\
\hline\\[-1.0pc] 
\hline
\end{tabular}} \label{tab:PX-PRL104-7-9-20-60}
\end{center}\vspace*{-0.5pc}
\end{table}\normalsize
\subsection{Conclusions}
\begin{figure}[!h]
\begin{center}
\raisebox{0pc}{\includegraphics[width=0.70\textwidth]{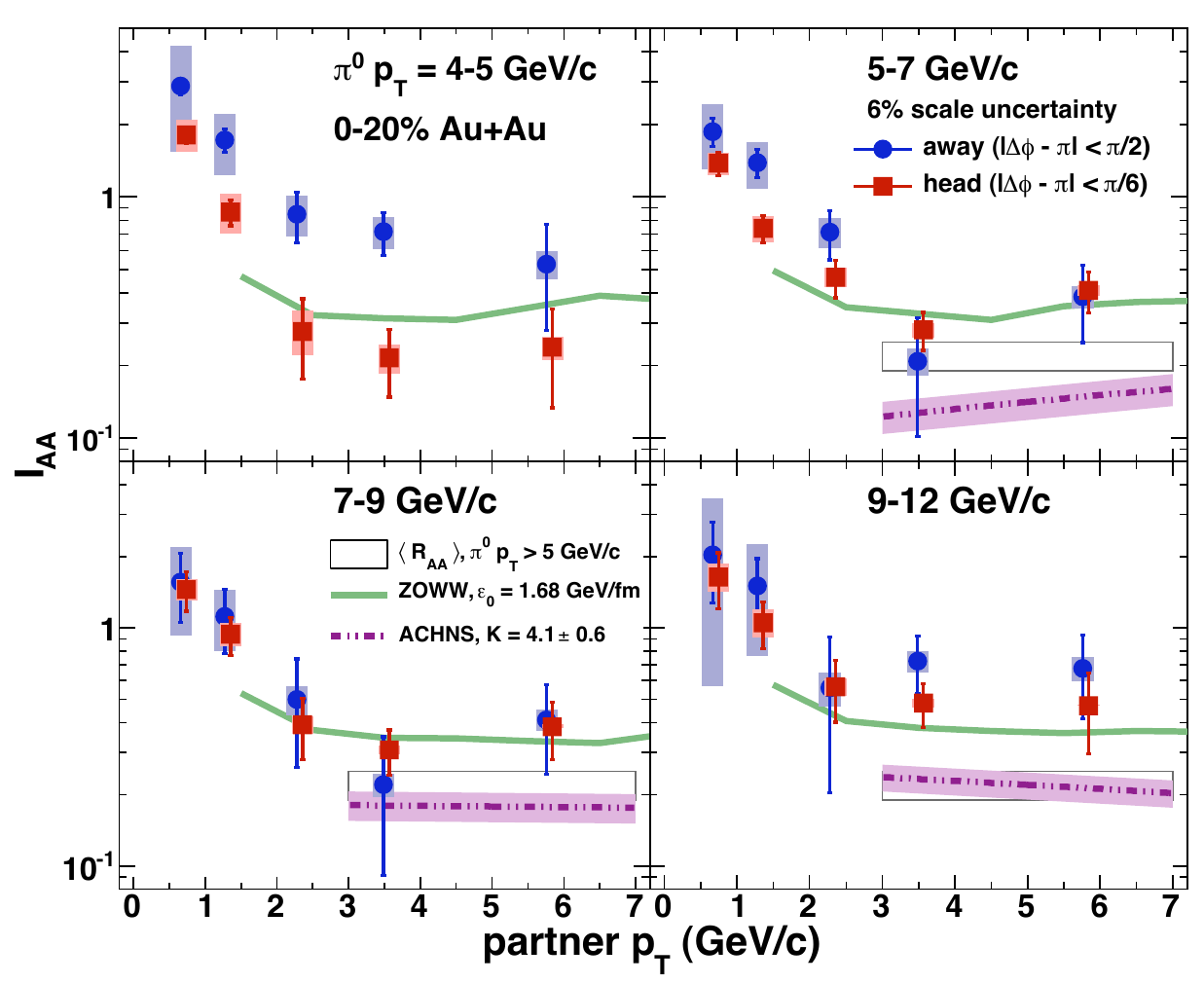}}\hspace*{0.2pc}
\end{center}\vspace*{-1.5pc}
\caption[]{\footnotesize PHENIX $I_{AA}$ distribution from [\Journal{PRL\ }{104}{252302}{2010}]}
\label{fig:IAAphenix}\vspace*{-0.5pc}
\end{figure}

\begin{figure}[!h]
\begin{center}
\raisebox{0pc}{\includegraphics[width=0.45\textwidth]{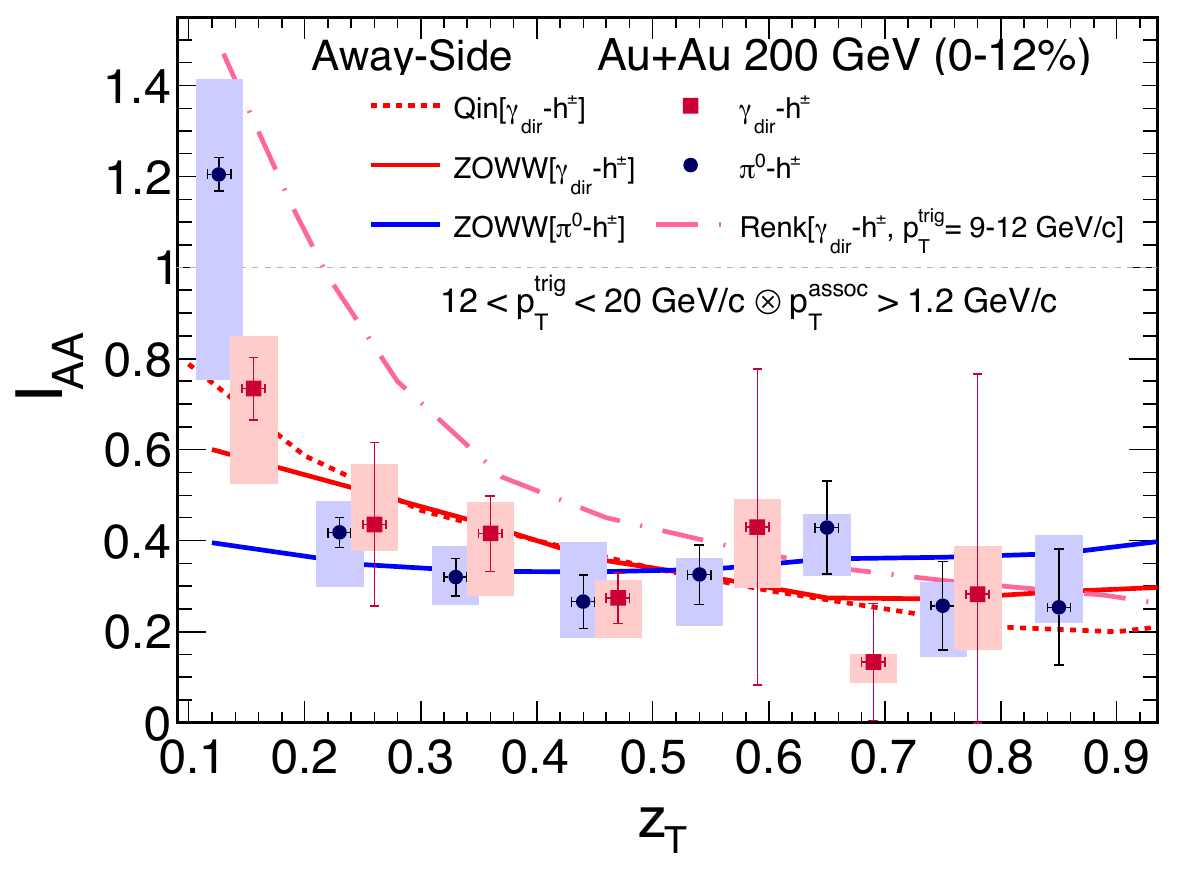}}\hspace*{0.2pc}
\raisebox{0pc}{\includegraphics[width=0.45\textwidth]{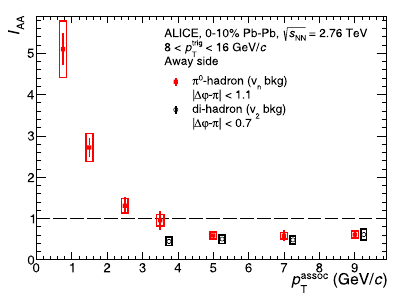}}\hspace*{0.2pc}
\end{center}\vspace*{-1.5pc}
\caption[]{\footnotesize (left) STAR $I_{AA}$ distribution from [\Journal{PLB\ }{760}{689}{2016}]; (right) ALICE $I_{AA}$ distribution from [\Journal{PLB\ }{763}{238}{2016}]}
\label{fig:IAAother}\vspace*{-0.5pc}
\end{figure}

It appears that the method works and gives consistent results for all the $\hat{q}L$ calculations 
shown (Tables \ref{tab:STARPLB771},\ref{tab:PX-PRL104-5-7-20-60},\ref{tab:PX-PRL104-7-9-20-60}). In the lowest $p_{Ta}\sim 1.5$ GeV/c bin the results are all consistent with the JET collaboration [\Journal{PRC\ }{90}{014909}{2014}] result, $\hat{q}=1.2 \pm 0.3$ GeV$^2$/fm or $\hat{q}L=8.4\pm 2.1$ GeV$^2$ for $L=7$ fm, the radius of an Au nucleus. However for $p_{Ta}> 2.0$ GeV/c all the results  are consistent  with $\hat{q}L=0$.  
Personally I think that this is where the first gluon emitted in the medium was inside the jet cone, so that all further emissions were also inside the jet cone due to the angular ordering of  \QCD\ so that there is no evident suppression; or that jets with fragments with $p_{T}\geq 3$ GeV/c,  which are distributed narrowly about the jet axis, are not strongly affected by the medium [arXiv:1302.2579]. I think that this also agrees with the observation in Fig.~\ref{fig:xTdist} that two or three orders of magnitude down in the $x_E=p_{Ta}/p_{Tt}$ distributions the A+A best fit is parallel to the p+p measurement which means that these A+A fragments are from jets that have not lost energy.  This is consistent with all the $I_{AA}=x_E^{AA}/x_E^{pp}=(p_{Ta}^{AA}/p_{Ta}^{pp})|_{p_{Tt}}$ distributions ever measured 
(e.g. Figs.~\ref{fig:IAAphenix}, \ref{fig:IAAother}) which decrease with increasing $p_{Ta}$ until $p_{Ta}\approx 3$ GeV/c and then remain constant because the A$+$A and p$+$p   distributions are parallel due to no jet energy loss for fragments in this range.

\end{document}